\documentclass[%
 aip,
 amsmath,amssymb,
 reprint,%
]{revtex4-1}

\usepackage{graphicx}
\usepackage{dcolumn}
\usepackage{bm}
\usepackage{tabularx}
\usepackage[utf8]{inputenc}
\usepackage[T1]{fontenc}
\usepackage{mathptmx}
\usepackage{etoolbox}
\usepackage{booktabs}
\usepackage{siunitx}
\usepackage{float}
\usepackage{easyReview}
\usepackage{makecell}
\usepackage[normalem]{ulem}
\usepackage{longtable}
\makeatletter
\def\@email#1#2{%
 \endgroup
 \patchcmd{\titleblock@produce}
  {\frontmatter@RRAPformat}
  {\frontmatter@RRAPformat{\produce@RRAP{*#1\href{mailto:#2}{#2}}}\frontmatter@RRAPformat}
  {}{}
}%
\makeatother
\begin{document}

\preprint{AIP/123-QED}

\title[]{High-throughput controlled droplets generation
 through a flow-focusing microchannel in shear-thinning fluids}
\author{Manohar Jammula}
\author{Somasekhara Goud Sontti}%
 \email{somasekhar.sonti@iitdh.ac.in}
\affiliation{ Multiphase Flow and Microfluidics (MFM) Laboratory, Department of Chemical Engineering, Indian Institute of Technology Dharwad, Dharwad, 580011, Karnataka, India.}


\begin{abstract}
\section*{ABSTRACT}
\noindent This study presents a three-dimensional transient computational fluid dynamics simulation of droplet generation in a flow-focusing microfluidic device using the Coupled Level Set and Volume of Fluid method for interface capturing. We systematically investigated the effects of shear-thinning non-Newtonian fluids, specifically carboxymethyl cellulose solutions, on the generation of mineral oil droplets. The rheological properties of carboxymethyl cellulose were analyzed at concentrations of 0.1\%, 0.25\%, 0.5\%, and 1.0\%, with the parameters $K$ and $n$ specified in terms of viscosity. The study examines the influence of rheological properties, continuous and dispersed phase flow rates, and interfacial tension forces on the generation and dynamics of controlled high-throughput droplets. The numerical results provide insights into droplet characteristics, including droplet length, velocity, formation frequency, liquid film thickness, pressure distribution, and flow regimes. The findings revealed that increasing the carboxymethyl cellulose concentration (0.1\%~to~1.0\%) and continuous phase flow rate ($10 ~\mu L/min $ to  $ 30 ~\mu L/min$) reduces droplet length by 53\% and 45\%, respectively. while increasing the dispersed phase flow rate ($2 ~\mu L/min $ to  $ 18 ~\mu L/min$) and interfacial tension results in  increased droplet length by 62\% and 42\%. At lower concentrations of carboxymethyl cellulose, plug-shaped droplets fully occupy the channel width in the squeezing regime. As concentration increases, a transition to the dripping and jetting regimes is observed. The study further explores non-dimensional parameters, expressing the Weber and modified Capillary numbers as functions of flow rate and interfacial tension. These insights impact drug delivery, material synthesis, and microfluidic technologies requiring precise droplet control. 
\end{abstract}

\maketitle

\section{\label{sec:level1}INTRODUCTION\\ }

In recent times, microfluidics emerged as a novel technology to generate microdroplets by controlling and manipulating fluid volumes at the micro level.\cite{gurkan2024next} 
The applications of microfluidics technology encompass a broad range of disciplines, including chemical processes, material synthesis, biological analysis, pharmaceuticals, food, and beverages.\cite{liu2020microfluidics, wang2024microfluidic, santana2018application, bandulasena2019droplet} Droplet microfluidics, a subcategory of microfluidics, involves manipulating and controlling the dispersed phase to generate droplets within an immiscible continuous phase typically driven by shear gradient.\cite{ren2015breakup} The monodispersed droplets generated via this technology exhibit a high surface-to-volume ratio, with their isolation from both the microchannel boundaries and adjacent droplets enabling their function as microreactors. \cite{sontti2019numerical,fatehifar2023numerical} This isolation significantly enhances mass and heat transfer within each droplet, facilitating more efficient chemical and biological reactions at the microscale.\cite{li2021heat, nieves2015openfoam} Droplet-based microfluidics has enabled a broad range of opportunities in biomedical applications, primarily in material synthesis, molecular biology analysis, cytology, disease identification, drug screening, cell culture, and tissue engineering. \cite{sackmann2014present, li2018microfluidic,chen2016controlled,jang2017droplet, yu2019microfluidic,chung2012microfluidic}\\

Droplets and bubbles in microfluidic devices are commonly generated through active or passive methods.\cite{liu2024breakup,liu2021electric, naji2023numerical} In the active droplet generation method, external energy is applied, such as thermal, electrical, acoustic waves, and magnetic forces to facilitate controlled droplet formation. \cite{chong2016active,kholardi2024understanding} In the passive method, immiscible continuous and dispersed phases are used. The shear force acting on the interfacial region on the dispersed phase and also the pressure gradient due to the continuous phase flow overcome the interfacial tension, eventually leading the dispersed phase to shred, followed by droplet formation.\cite{svetlov2021mathematical} Thus, it is important to consider fluid properties and operating conditions for droplet generation by a passive method. Nevertheless, the design of the microfluidic device serves an equally critical role. The most widely used passive microfluidic configurations include co-flow, cross-flow junction, and flow-focusing device.\cite{teh2008droplet}Among these microfluidic configurations, the flow-focusing device is simple and has the merit of accurate control over the droplet formation and size with high throughput.\cite{lashkaripour2019performance}\\

Over the years, significant progress has been achieved in the fabrication technology of microfluidic devices. Some techniques include chemical process, including wet and dry etching, soft lithography, photolithography, micromilling, and 3D printing. \cite{niculescu2021fabrication} These advancements have facilitated more in-depth studies of microfluidic devices. In particular, detailed investigations on droplet formation in flow-focusing microfluidic devices are crucial, as a comprehensive understanding of these processes is essential due to their notable advantages and broad applications across various fields. The geometry configuration of a flow-focusing device significantly affects the droplet formation and consequent droplet properties such as size, shape, and frequency. \cite{ngo2016effects} \citet{mastiani2017flow} experimentally investigated the impact of flow-focusing device modification with an angle at the junction of the microchannel. The considered angles at the junctions were $30^\circ$ and $90^\circ$. It was found that the measured droplet size in the dripping regime was found to be smaller at $30^\circ$, while in the jetting regime, the droplet size was smaller at $90^\circ$. However, the droplet formation frequency was higher at $90^\circ$ due to the squeezing force acting on the dispersed phase being perpendicular and maximum. Furthermore, the operating conditions, such as the flow rates of continuous and dispersed phases, also play an important role in droplet formation.\\
 
 \citet{costa2017studies} experimentally investigated the droplet formation in a flow-focusing microchannel to study flow regimes with the effects of flow rates. Three different regimes observed during the experiments are squeezing, dripping, and jetting. Notably, squeezing was found when flow rate ratios were between 0.5 and 1, whereas dripping was found when flow rate ratios were lower. Considering practical applications, it was concluded that the dripping regime is more favorable as smaller droplets were created in this regime compared to the squeezing regime. In addition, a correlation as a function of capillary number is developed to estimate the droplet diameter. \citet{liu2024breakup} examined the impact of continuous phase velocity in a flow-focusing (400$\times$400 $\mu$m) microchannel. It was reported that with increasing flow rates in the continuous phase, the curvature of the droplets decreased, indicating that the opening angle of the inner concave interface (normalized) increases with continuous phase; hence, neck break-up becomes a self-similar process independent of breakup dynamics. The results indicated that in the slow squeezing stage, the dimensionless neck width of the droplet is linear to time, while in rapid squeezing and rapid pinch-off stages, it followed a power-law relation with time. \citet{qian2022determination} proposed a novel approach to find the droplet velocity and film thickness for droplets generated in a microfluidic flow-focusing configuration. The experimental results of three different fluid combinations showed excellent agreement with the proposed approach, with an error of less than 4\%. \citet{wu2019flow} conducted experiments using three different flow-focusing square microchannels. The study focused on visualizing the dynamics of the dispersed phase and formation of droplet slugs at the channel junction and further downstream in the main microchannel. The non-dimensional droplet length was characterized using the Capillary number and Weber number as functions of the continuous and dispersed phase velocities under various operating conditions.

A considerable amount of work on microfluidics has been reported over the past two decades using Newtonian fluids, which have constant viscosity regardless of shear rate. Droplet generation using a Newtonian fluid, the interfacial dynamics primarily depend on the constant viscosity and interfacial effects\cite{wu2017liquid, liu2024breakup}, and droplet size and flow conditions such as slug flow, squeezing, jetting, and stratified flow are investigated by altering the flow rates of both the phases\cite{guan2019liquid, kurniawan2023formation}. In contrast, the viscosity of non-Newtonian fluids varies with shear rate and exhibits complex rheological behavior. Some of the non-Newtonian fluid microfluidic systems are yield stress\cite{shang2023formation}, viscoelastic\cite{du2016breakup}, and shear thinning\cite{fu2016breakup}. In practical applications, the fluids found in food processing, chemical process, biological analysis, and material synthesis \cite{guo2020continuous, song2006chip}  behave non-Newtonian in nature and exhibit non-homogeneity. \citet{chen2021pressure} studied droplet formation in a flow-focusing microfluidic device by using Newtonian and non-Newtonian fluids as a primary phase. During droplet formation, a thin layer of the continuous phase was observed between the droplet and the channel wall when a non-Newtonian fluid was used, attributed to its higher apparent viscosity. For the non-Newtonian continuous phase, the droplet formation consistently occurred in the dripping flow pattern across all tested flow rates. In contrast, for Newtonian fluids, both squeezing and dripping flow patterns were observed depending on the flow conditions. \citet{pan2021flow} reported droplet formation using both Newtonian and non-Newtonian dispersed phases. Increasing the concentration of polystyrene in the fluorobenzene solution enhanced the non-Newtonian shear-thinning behavior. As the concentration of the dispersed phase increased, the formation of satellite droplets became more prominent compared to the Newtonian case.

\citet{xue2019non} reported the generation of non-Newtonian droplets in a flow-focusing microchannel. The experiment's major observation was that as the concentration of the disperse phase increased, the flow regime shifted from the dripping stage to the jetting stage, and satellite droplets were observed at the tail of the droplet generation. \citet{du2018breakup} experimentally investigated shear-thinning of the droplet formation in a flow-focusing microchannel. The dynamics of droplet growth and squeezing are dominated by squeezing pressure, causing the dispersed thread to thin faster as the flow rate increases. In the droplet growth stage, viscous stresses become prominent, and during the squeezing stage, they delay the rupture of the dispersed thread. A recent study of \citet{rostami2018generation}  established the formation of Newtonian and Non-Newtonian droplets through a cross-junction microchannel. The silicone oil and xanthan gum solutions (0.2, 0.3, and 0.5 wt\%) with surfactant were used as the continuous phase and dispersed phase. With an increased flow rate ratio and capillary number, when the secondary phase is pure water, mono-dispersed droplets are observed. In contrast, when the xantham solution with surfactant is used, poly-dispersed droplets are observed as well as satellite droplets.

Computational fluid dynamics (CFD) is a powerful technique for developing such models and a deeper understanding of droplet formation in complex fluids and quantifying the key physical parameters such as pressure, velocity, viscosity, droplet size, shape, and frequency. To model the droplet formation in an immiscible multiphase flow, interface capturing is a pivotal criterion. The most commonly used interface capturing techniques are the phase field method~\cite{bariki2022flow}, Lattice Boltzmann method (LBM)\cite{wang2023numerical}, volume-of-fluid (VOF) method~\cite{hirt1981volume,zhang2022effect}, and level-set (LS) method~\cite{osher1988fronts,howard2021conservative}. In a recent study, \citet{chao2021cfd} used the VOF method to predict the droplet's internal circulation. It was analyzed from the topology of internal circulation that the droplet movement was majorly contributed by the pressure and surface tension forces over viscous forces. With increasing viscosity ratio between the disperse and continuous phases, internal circulation became smaller due to the shear rate. With vertical inlet configurations, the squeezing regime's mixing efficiency was larger than in the shrinking and stabilized stages for viscosity ratios 1 to 3.83.  The mixing efficiency was also increased by 14.2 $\% $ by adding vertical continuous phase flow inlet configurations.\\

\begin{figure*}[!ht] 
	\centering
	\includegraphics[width=1.0\textwidth]{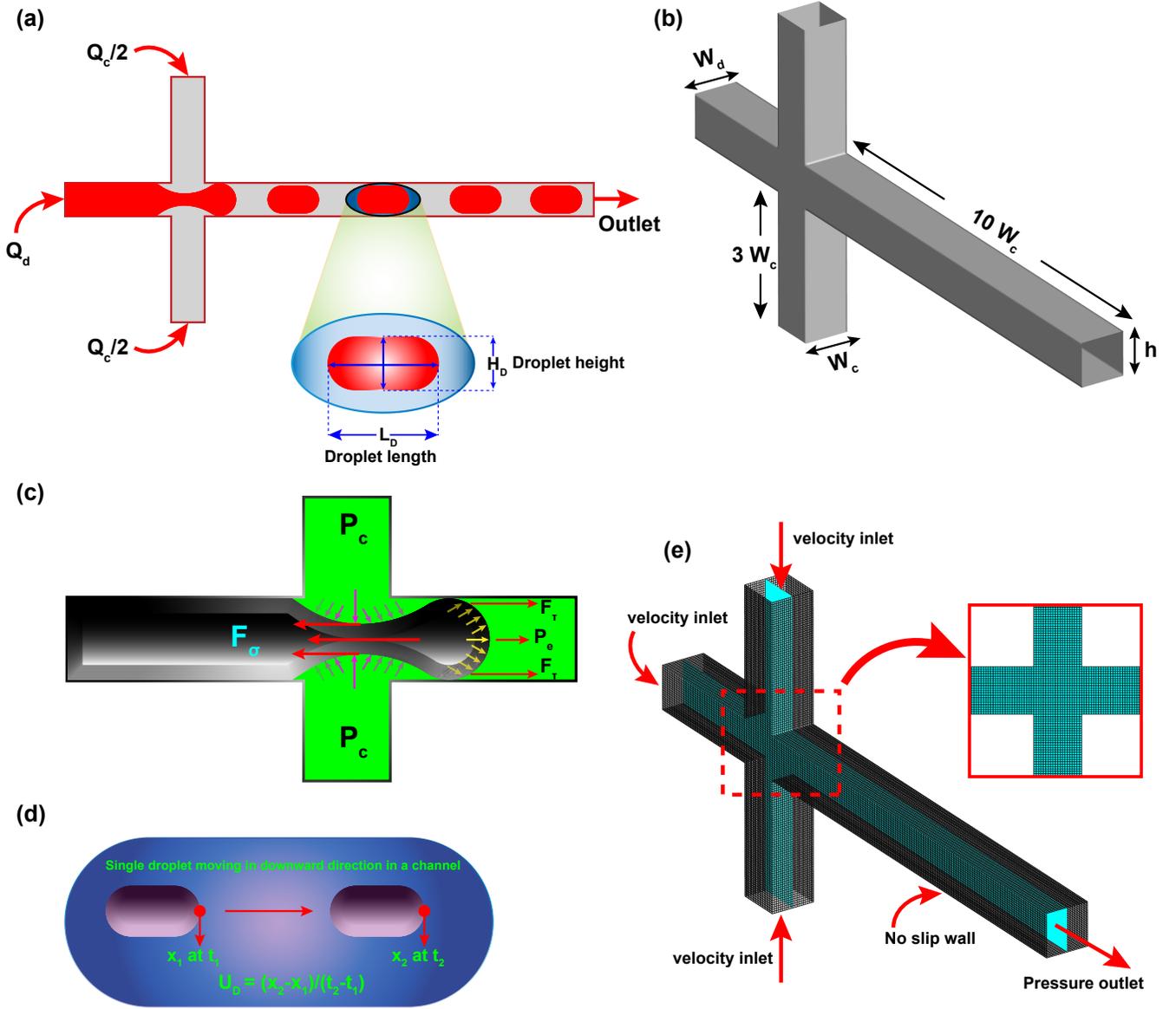} 
	\caption{ (a) Schematic illustration of droplet formation in a square flow-focusing microfluidic device where the droplet length from the rear to the nose is denoted with $L_{D}$ and height is denoted $H_{D}$, (b) three-dimensional square flow-focusing microchannel with channel dimensions, (c) illustration of the typical forces exerted on the dispersed phase upon droplet generation in a flow-focusing microchannel, (d) droplet velocity measurement, and (e) Structured mesh of the 3D domain and all the boundary conditions.}
	\label{fig-1}
\end{figure*}


\citet{fatehifar2021non} worked on numerical investigation of non-Newtonian droplets surrounded by a Newtonian fluid using the VOF method in a flow-focusing microchannel. There was an observed transition from dripping to tip-streaming with increased shear-thinning fluid concentration. The VOF model performs well in conserving mass in the most accurate way, however the step discontinuity makes it difficult to capture the curvature at interfaces using the volume fraction function. Interestingly, the interface curvature capturing is smoothly computed with the LS method. Therefore, combining VOF and LS methods, coupled level set and volume-of-fluid (CLSVOF), allows for smooth interface capturing while maintaining the highest mass conservation accuracy. \citet{sontti2019understanding} investigated bubble formation in a shear-thinning fluid with VOF and CLSVOF, and the bubble interface using CLSVOF was smooth compared to the VOF method. \citet{carneiro2019high} computationally compared droplet generation in a flow-focusing microfluidic device employing VOF and CLSVOF methods. The droplet shape from the CLSVOF modeling was found to be identical to the experimental droplet shape. Further, in a recent study \citet{sontti2023regulation} numerically investigated the droplet formation in a flow-focusing device using the CLSVOF method. The effect of geometrical modifications on droplet formation and their dynamics was reported. A study by \citet{wang2018numerical} computationally investigated the droplet coalescence with advancing fluid meniscus in a microchannel using a CLSVOF. The results are in excellent concordance with experimental data.\\

Recent works have successfully implemented CLSVOF for non-Newtonian systems. \citet{ohta2019three} implemented the CLSVOF model to capture the bubble deformation surrounded by a liquid that possesses a non-Newtonian behavior of the hybrid model, a combination of the shear-thinning Carreau model and viscoelastic model in a liquid column. The deformation of the bubble shapes, such as oblate, cap, and sharp cusp, is compared with a Newtonian fluid. \citet{du2020numerical} computationally evaluated bubble generation in a flow-focusing microchannel, where the bubbles were surrounded by a shear-thinning power-law fluid. Using the CLSVOF interface-capturing method, they simulated the subsequent breakup of the bubble into four branches. The model effectively captured the formation of four sub-bubbles, each having lengths equal to their corresponding symmetric counterparts. Beyond microfluidics, CLSVOF has also been successfully applied in other multiphase flow problems involving interfacial dynamics. \citet{bao2025dynamic} used CLSVOF to model droplet interaction on a liquid film, capturing the interface evolution and splashing behavior. \citet{li2024numerical} extended the method to simulate $SO_2$ absorption into droplets and mass transfer across deformable interfaces in reactive gas–liquid systems. The CLSVOF method is implemented in multiphase systems such as droplet breakup ~\cite{mora2024taylor}, droplet falling in a liquid pool ~\cite{singh2023hanging}, and droplet evaporation.\cite{xia2024improved}

Despite extensive research on droplet generation in microchannels using CFD simulations, most studies have primarily focused on fluids obeying Newtonian behaviour. While the role of non-Newtonian fluids and fluid concentration, dependent on rheological parameters of shear-thinning fluids, on droplet formation and dynamics is less explored. Fundamental analysis of passive droplet generation involving non-Newtonian fluids is still required to gain deeper insights into their influence on droplet formation, flow behavior, interfacial dynamics, and flow regime transition in microfluidic systems. This analysis is essential for optimizing the processes, particularly in applications involving complex fluids, such as mass transfer and chemical reactions\cite{madadelahi2017droplet}, drug delivery\cite{shang2017emerging}, and metal extraction.\cite{fernandez2024high}\\

The main objective of this study is to conduct a comprehensive numerical analysis using a CFD model with the interface capturing through the CLSVOF method to investigate the impact of the non-Newtonian shear-thinning behavior on droplet formation. The numerical analysis of droplet dynamics includes both quantitative and qualitative aspects of key parameters such as droplet size (length), velocity, formation frequency, flow regime transitions, liquid film thickness, dynamic viscosity, and pressure distribution. In addition, the flow regime maps specifically for the non-Newtonian systems will be developed to identify the droplet breakup regimes. \\

This work is structured as follows. In Sec. II, we outline the problem formulation, the configuration of the device, and the continuous and dispersed phase fluid properties. Section III presents the governing equations and the details of the interface capturing methodology. In Sec. IV, we outline the solver settings, CFD model validation, and grid independence analysis. The results and discussion are provided in Sec. V, where we examine the effects of continuous-phase concentration, flow rate ratio, and interfacial tension on droplet formation, along with flow regime maps. Finally, Sec. VI summarizes the key findings of this study.

\section{PROBLEM FORMULATION}

\begin{figure*}[!ht] 
	\centering
	\includegraphics[width=\textwidth]{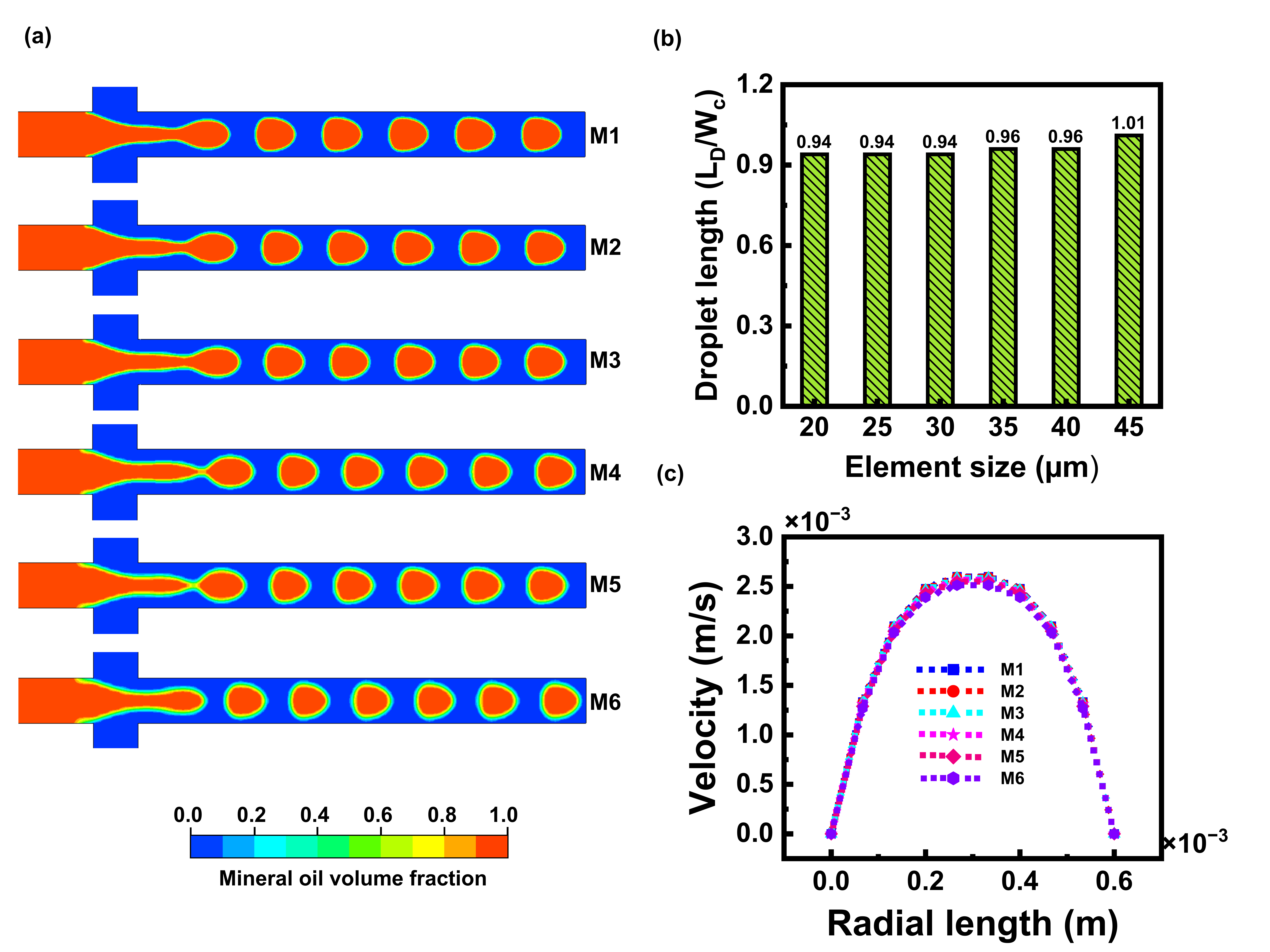} 
	\caption{ Grid independence study: Comparison of (a) mineral oil droplet volume fraction for different mesh elements, (b) mineral oil non-dimensional droplet length, and (c) radial velocity for CMC- 1\% at $Q_{c} = 20 ~\mu L/min$ and $Q_{d} = 10 ~\mu L/min$.}
	\label{fig-3}
\end{figure*}

  The droplet formation in a three-dimensional flow-focusing microfluidic device and the magnified view of the droplet with droplet length and height, are denoted with $L_{D}$ and $H_{D}$, respectively shown in Fig.\ref{fig-1}a. The standard square flow-focusing microfluidic device with a channel width of continuous and dispersed phase inlets and height $(W_{c} =W_{d} =h)$ is 600 $\mu m$. The channel entrance length of the continuous phase and dispersed phase are three times the channel width (i.e., 1800 $\mu m$), while the main channel length is ten times the width of the channel (i.e., 6000 $\mu m$) as shown in Fig. \ref{fig-1}b.  In this work, the dispersed phase (mineral oil) is Newtonian fluid that enters the device from the main inlet of the channel, whereas the continuous phase ( sodium carboxymethyl cellulose (CMC)) is a shear-thinning non-Newtonian fluid enters from the two side inlets into the device and the continuous phase obeys the shear-thinning power-law model ($\mu=K \dot{\gamma}^{\mathrm{n}-1}$)\cite{chen2020modeling}, where $K$ is consistency index, $n$ is a power law index and $\dot{\gamma}$ is a shear rate. The fluid properties used in this work are shown in Table \ref{tab:1}.
 The comparison of the simulated dynamic viscosity of CMC solutions with the experimental data as a function of shear rate is shown in Fig. S1, and the results match the experimental data with less than 5\% error.

The droplet formation in a flow-focusing microfluidic device occurs by a passive mechanism. The dispersed phase passes through the main inlet, and both the phases are in contact at the channel junction. The entered dispersed phase blocks the complete channel width and blocks the continuous phase to flow. As a result, pressure builds up upstream of the continuous phase. A combined force of inertial and shear exerts hydrodynamic pressure on the dispersed phase.\cite{fu2016breakup} This pressure causes the dispersed phase to squeeze into the main channel, leading to the evolution of neck formation and eventual neck break up and droplet formation. The primary forces involved during the passive droplet generation along with the direction of forces are illustrated in Fig. \ref{fig-1}c, where the $P_{c}$ is the hydrodynamic pressure due to the inertial and viscous forces of the continuous phase, $F_{\sigma}$ is the interfacial tension force, $P_{e}$ is the downstream pressure, and $F_{T}$ is the shear force. Fig. \ref{fig-1}d illustrates the typical procedure to quantify the droplet velocity of an independent single and stable droplet moving in a downward direction with respect to time, where $X_{1}$ and $X_{2}$ are the position of the droplet front end at corresponding time $t_{1}$ and $t_{2}$, respectively.

\section{MATHEMATICAL MODELING}
In this study, the continuous and dispersed fluids are immiscible, isothermal, and incompressible, with the flow exhibiting a laminar pattern. The emergence of droplets and the analysis of droplet dynamics are performed computationally by discretizing a single set of mass and momentum conservation equations, given as follows:\\

Conservation of mass equation:
\begin{equation}
\frac{\partial \rho}{\partial t}+\nabla(\rho \vec{u})=0
\end{equation}
Conservation of momentum equation:
\begin{equation}
\frac{\partial}{\partial t}(\rho \vec{u})+\nabla \cdot(\rho \vec{u} \vec{u})=-\nabla p+\nabla \cdot\tau+\rho \vec{g}  +\vec{F_\sigma}
\end{equation}

\begin{equation}
    \tau = \mu(\dot{\gamma})\left(\nabla \vec{u}+\nabla \vec{u}^T\right)
\end{equation}
for shear thinning power-law fluid 
\begin{equation}
    \mu(\dot{\gamma})  =K \dot{\gamma}^{n-1}
\end{equation}
Here, $\vec{u}$ represents the fluid velocity, $\rho$ is the density, $\mu$ denotes the dynamic viscosity, and $p$ is the pressure. The terms $\vec{g}$ and $\vec{F_\sigma}$ correspond to the gravitational force and the force due to interfacial tension, respectively. $K$ and $n$ denote the consistency and power-law indices, respectively, while $\dot{\gamma}$ represents the local shear rate, calculated from the velocity gradient tensor.
\begin{equation}
\dot{\gamma}=\sqrt{2(\vec{D}: \vec{D})}
\end{equation}
Where $\vec{D}$ is defined as a rate of deformation tensor
\begin{equation}
\vec{D}=\left(\nabla \vec{u}+\nabla \vec{u}^T\right)
\end{equation}
The Bond number ($B o=\Delta \rho g r^2 / \sigma$) is less than one, indicating that gravitational effects are negligible in the momentum equation. However, the interfacial tension force, which plays a crucial role in immiscible and capillary flows, is explicitly considered in the momentum equation.

\subsection{Interface capturing methodology}

\noindent The key feature of the CLSVOF technique is its ability to compute the interface's normal curvature using a smooth level set (LS) function. The level set function ($\Phi$) is a signed distance function that depends on space ($\vec{\zeta}$) and time (t). The fluid phase is distinguished depending on the sign variations of ($\Phi$) across the interface. The tracking of interface movement is solved explicitly by advecting the volume fraction and level set. The transport equations for the level set and volume fraction are expressed as follows:

\begin{equation}
\frac{\partial \Phi}{\partial t}+\vec{u} \cdot \nabla \Phi=0
\end{equation}

\begin{equation}
\frac{\partial \alpha}{\partial t}+\vec{u} \cdot \nabla \alpha=0
\end{equation}

Where $\vec{u}$ is the velocity and $\Phi$ is the level set function, and $\alpha$ is the volume fraction of the fluid ranging from 0 to 1 in a given discretized unit cell. A level set function measures the positive and negative magnitude of $\gamma$, where $\gamma$ = $\gamma$( $\vec{\zeta}$)  is the possible nearest distance from the interface to the space ($\vec{\zeta}$) in the bulk fluid at time t.

\begin{equation}
\Phi(\vec{\zeta}, t)= \begin{cases}\gamma & \text { if } \zeta \text { is completely in the mineral oil } 
\\ 0 & \text { if } \zeta \text { is in the interface between two phases } 
\\ -\gamma & \text { if } \zeta \text { is completely in the CMC phase }\end{cases}
\end{equation}
For a continuous variation of the interface, a piecewise linear interface-capturing (PLIC) scheme is employed to reconstruct the interface in each cell for the succeeding time interval.\cite{holt2012numerical} 
The viscosity field is updated at every iteration based on the shear rate. The viscosity field and density variations across the interface are continuously computed from the Heaviside function. \cite{deka2019coalescence} The viscosity and interface are coupled using a Heaviside function in the CLSVOF modeling.  

\begin{equation}
\rho(\Phi)=H(\Phi) \rho_d+(1-H(\Phi)) \rho_c ,
\end{equation}
\begin{equation}
\mu(\Phi)=H(\Phi) \mu_d+(1-H(\Phi)) \mu_c 
\end{equation}

The smoothed Heaviside function is expressed as such:
\begin{equation}
H(\Phi)= \begin{cases}0 & \text { if } \Phi<-\omega \\ \frac{1}{2}\left[1+\frac{\Phi}{\omega}+\frac{1}{\pi} \sin \left(\frac{\pi \Phi}{\omega}\right)\right] & \text { if }|\Phi| \leq \omega \\ 1 & \text { if } \Phi>\omega\end{cases}
\end{equation}

Where $\omega$ is the thickness of the interface.\\

The interfacial tension force $\vec{F}$ in the momentum equation  is determined by the continuum surface force (CSF) model \cite{brackbill1992continuum} as shown below:
\begin{equation}
\vec{F}=\sigma \kappa(\Phi) \delta(\Phi) \nabla \Phi,
\end{equation}
Where $\sigma$, $\kappa(\Phi)$, and $\delta(\Phi)$ are the interfacial tension, interface curvature, and smoothed Dirac delta function.
 
\begin{equation}
\kappa(\Phi)=\nabla \cdot \frac{\nabla \Phi}{|\nabla \Phi|}
\end{equation}

\begin{equation}
\delta(\Phi)=\left\{\begin{array}{ll}
0 & \text { if }|\Phi| \geq \omega \\
\frac{1}{2 \omega}\left(1+\cos \left(\frac{\pi \Phi}{\omega}\right)\right) & \text { if }|\Phi|<\omega
\end{array} \right.
\end{equation}

The CLSVOF method integrates the VOF, and Level-Set approaches from  Eq. 1-15 by establishing an algebraic relationship between them, ensuring mass conservation while maintaining a smooth and continuous interface. The flow chart of the CLSVOF algorithm is shown in Fig. S2\\

\begin{table}[h!]
\caption{\label{tab:1} Physical properties of the fluid used in CFD simulations \cite{du2018breakup}.}
\centering
\small 
\begin{tabular}{lcccc}  
\toprule
\textbf{Fluid} & \textbf{$\rho$(\si{kg/m^3})} & \textbf{$\mu_{c}$ or $K$ (\si{Pa.s^n})} & \textbf{$n$} & \textbf{$\sigma$ (\si{mN/m})} \\
\midrule
Mineral oil & 875.1 & 0.026 & 1.0 & - \\
CMC-0.1\% & 998.9 & 0.0257 & 0.865 & 2.9 \\
CMC-0.25\% & 999.5 & 0.0489 & 0.830 & 2.8 \\
CMC-0.5\% & 1000.5 & 0.1485 & 0.792 & 2.9 \\
CMC-1.0\% & 1002.8 & 0.9671 & 0.696 & 2.9 \\
\bottomrule
\end{tabular}
\end{table}

\section{SOLUTION METHODOLOGY  AND MODEL VALIDATION   }

\subsection{Solver settings and solution methodology}
The finite volume method (FVM) is employed to discretize the aforementioned transient partial differential equations 1-15. Pressure-velocity coupling is implemented using pressure implicit splitting operators (PISOs) in incompressible flows to ensure mass conservation. The momentum equation and level set function are discretized by the second-order upwind method and first-order upwind method, respectively. The pressure staggering option (PRESTO) scheme is used to compute the pressure values at each control volume. The piecewise linear interface construction approach is used to solve the volume fraction. The adaptive time step is considered to solve the governing equations with an initial time step $10^{-3}$ with convergence criteria $10^{-4}$ and Courant number~(Co) less than 0.25. No slip boundary condition is specified to the wall of the microchannel, and a pressure outlet boundary condition is specified. The inlet conditions for both continuous and dispersed phases specified are constant velocity inlet boundary conditions, shown in Fig. \ref{fig-1}e. Initially, it is assumed the microchannel is completely filled with a continuous phase before the dispersed phase is entered.  In this study, a contact angle of $180^{\circ}$ was employed to represent a perfectly non-wetting condition between the fluid interface and the channel walls. This condition ensures that the continuous phase fully wets the channel walls, effectively isolating the droplets and preventing interfacial distortion near the boundaries\cite{filimonov2021toward}.In practical applications, such as mass transfer and reaction kinetics between droplets and surrounding fluids, maintaining droplet isolation from the walls increases the specific interfacial area available for transport processes. Consequently, the mass transfer coefficient is enhanced, improving system performance. \cite{nieves2015openfoam} The list of solver settings and numerical schemes used in the present study are reported in the Table \ref{tab:Solversettings}.\\

All simulations were performed on a high-end workstation equipped with an Intel i9 processor, 64 GB of RAM, and 24 cores. The computational time required for each simulation was approximately 33 hours.
\begin{table}
\centering
	\caption{List of CLSVOF model solver settings and discretization schemes used in the present study.}
	\vspace{0.1cm}
	\label{tab:Solversettings}
         \renewcommand{\arraystretch}{0.70}
	\resizebox{0.4\textwidth}{!}{
		\begin{tabular}{ll}
			\specialrule{.1em}{.05em}{.05em} 
			\multicolumn{1}{c}{Model} & 
            \multicolumn{1}{c}{Scheme} \\ \specialrule{.1em}{.05em}{.05em} 
			Multiphase model & eulerian\\
            Interface capturing & CLSVOF\\
            Viscous model & laminar\\
			Pressure\textendash velocity coupling & PISO\\
			Pressure & PRESTO\\
			Momentum  & second order upwind \\
            Volume fraction  & first order upwind\\
            Level set function & first order upwind\\
			Flow Time  & 10 s\\
			Time scheme & adaptive\\
            Global Courant number & 0.25\\
			Inlet & constant velocity\\
            Wall & no-slip condition\\
            Outlet & pressure outlet\\
             Contact angle & $180^{\circ}$\\

			\specialrule{.1em}{.05em}{.05em} 
		\end{tabular}%
	}
\end{table}

\subsection{Grid independence analysis}
We systematically performed a grid-independent analysis with six different mesh element sizes of 20 $\mu m$, 25 $\mu m$, 30 $\mu m$, 35 $\mu m$, 40 $\mu m$, and 45 $\mu m$  named M1, M2, M3, M4, M5, and M6, respectively, with hexahedral structured mesh elements (shown in Table \ref{tab:2}). As the mesh element size increased, the total number of mesh elements decreased, and the corresponding mesh elements from M1 to M6 are 520000, 276000, 160000, 95000, 66000, and 44000   elements, respectively. Figure \ref{fig-1}e shows the structured mesh of M3, highlighting the cross-sectional view of the mesh on the mid-plane of the microfluidic device.  \\

The interface resolution of all the six mesh sizes is shown in Fig. \ref{fig-3}a, as the mesh element size increased from 20 $\mu m$ to 45 $\mu m$, the interface thickness changed from the fine interface to the coarse interface. The corresponding non-dimensional droplet length with respect to mesh elements is shown in Fig. \ref{fig-3}b. The droplet length decreases with the increase in the number of mesh elements, and the variation of droplet length from M1 to M3 is negligible. The radial velocity for all the mesh elements is shown in Fig. \ref{fig-3}c. For all the mesh elements, the velocity profile is parabolic and identical. However, the velocity profiles for M1, M2, and M3 are almost identical. Considering the droplet size variation, interface resolution, and velocity profile, the mesh M3 results are identical to the finer mesh M1. Therefore, further investigation of droplet formation and the effects of different parameters on droplet dynamics are continued with mesh M3.
\subsection{Model validation}

\begin{figure*}[!ht] 
	\centering
	\includegraphics[width=\textwidth]{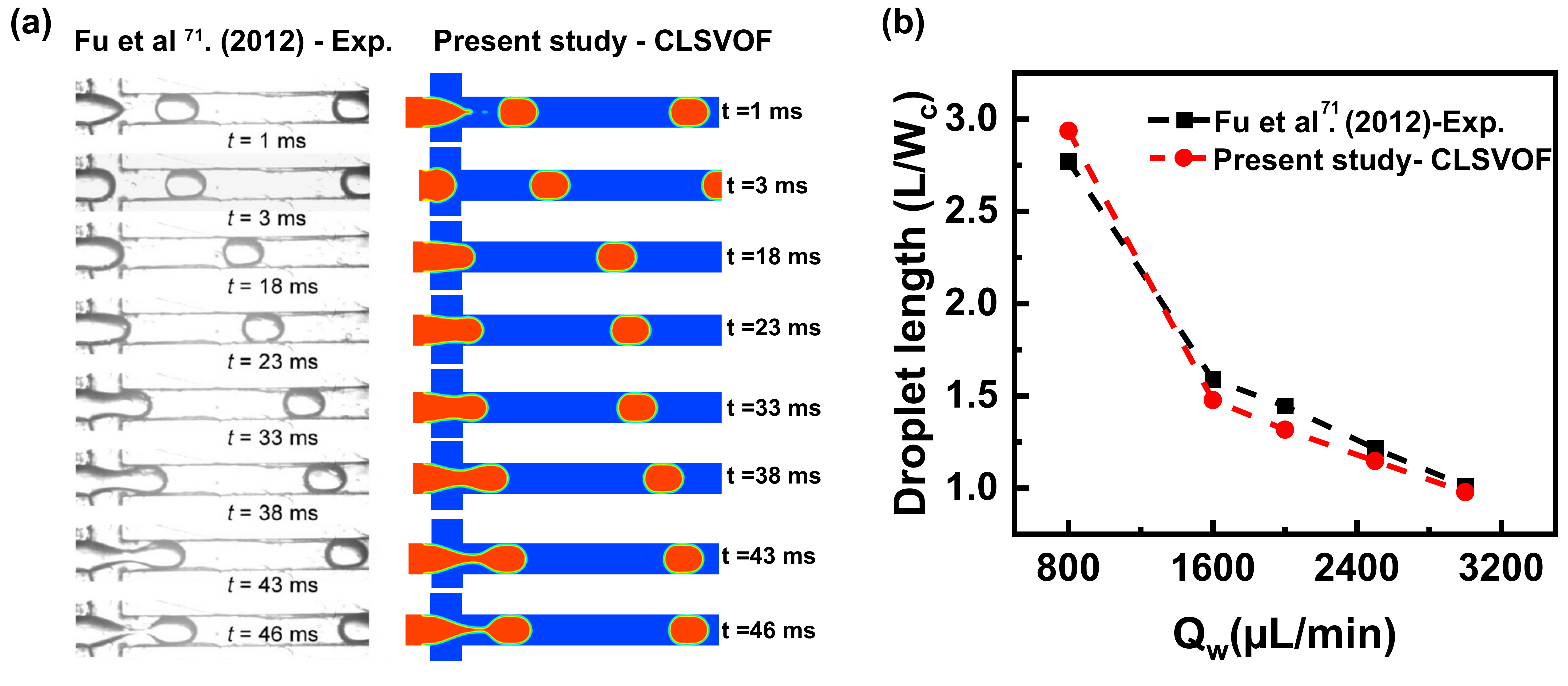} 
\caption{ (a) Comparison of the temporal variation of oil droplet formation with experimental findings of \citet{fu2012droplet} at $Q_{o} = 300~\mu L/\text{min}$ and $Q_{w} = 2000~\mu L/\text{min}$, Reproduced with permission from Fu et al., Chem.
Eng. Sci. 84, 207–217 (2012). Copyright 2012 Elsevier and (b) Comparison of CLSVOF model predicted non-dimensional droplet length with 
 the experimental findings of \citet{fu2012droplet} for  the variations in the continuous phase flow rate at a fixed $Q_{o} = 800~\mu L/\text{min}$.}
	\label{fig-2}
\end{figure*}

First, it is necessary to validate CFD model data to analyze the efficiency of the aforementioned CLSVOF model. The CLSVOF model results are validated with \citet{fu2012droplet} experimental work on droplet formation and length both qualitatively and quantitatively. The qualitative validation is the oil droplet formation mechanism at the oil and water flow rates 300 $\mu L/min$ and 2000 $\mu L/min$, respectively. The temporal variation of oil droplet formation in a dripping regime is compared with the CLSVOF model, as shown in Fig. \ref{fig-2}a. It is apparent that the CLSVOF method captures the droplet formation mechanism and flow regime for the experimental operating conditions considered. Further, the effect of droplet length with continuous phase flow rate ($Q_{w} = 800  ~to  ~3000 ~\mu L/min$) at a fixed oil flow rate ($Q_{o} = 800 ~ \mu L/min$) is also compared. The oil droplet length is represented in a non-dimensional length (L/$W_{c}$) as shown in Fig. \ref{fig-2}b, and the predicted results showed excellent concordance with the experimental observations. Therefore, the CFD model adequately describe flow physics, both qualitatively and quantitatively, in agreement with literature data.
 
\begin{table}[h!]
\caption{\label{tab:2} Hexahedral structured grid study with different mesh element sizes and droplet length comparison.}
\centering
\begin{tabular}{lccc}  
\toprule
\textbf{Mesh} & \textbf{ size ($\mu m$)} & \textbf{Total elements} & \textbf{Droplet length ($L_{D}/W_{c}$)} \\
\midrule

M1 & 20 & 520000 & 0.94 \\
M2 & 25 & 276000 & 0.94 \\
M3 & 30 & 160000 & 0.94 \\
M4 & 35 & 95000 & 0.96 \\
M5 & 40 & 66000& 0.96 \\
M6 & 45 & 44000 & 1.01\\
\bottomrule
\end{tabular}
\end{table}

\section{RESULTS AND DISCUSSION}
The effects of fluid properties and operating conditions on droplet formation are discussed in the following subsections, including the influence of continuous phase concentration, the flow rates of both the continuous and dispersed phases, and interfacial tension forces.

\subsection{Effect of continuous phase concentration}

\begin{figure*} 
	\centering
	\includegraphics[width=0.80\textwidth]{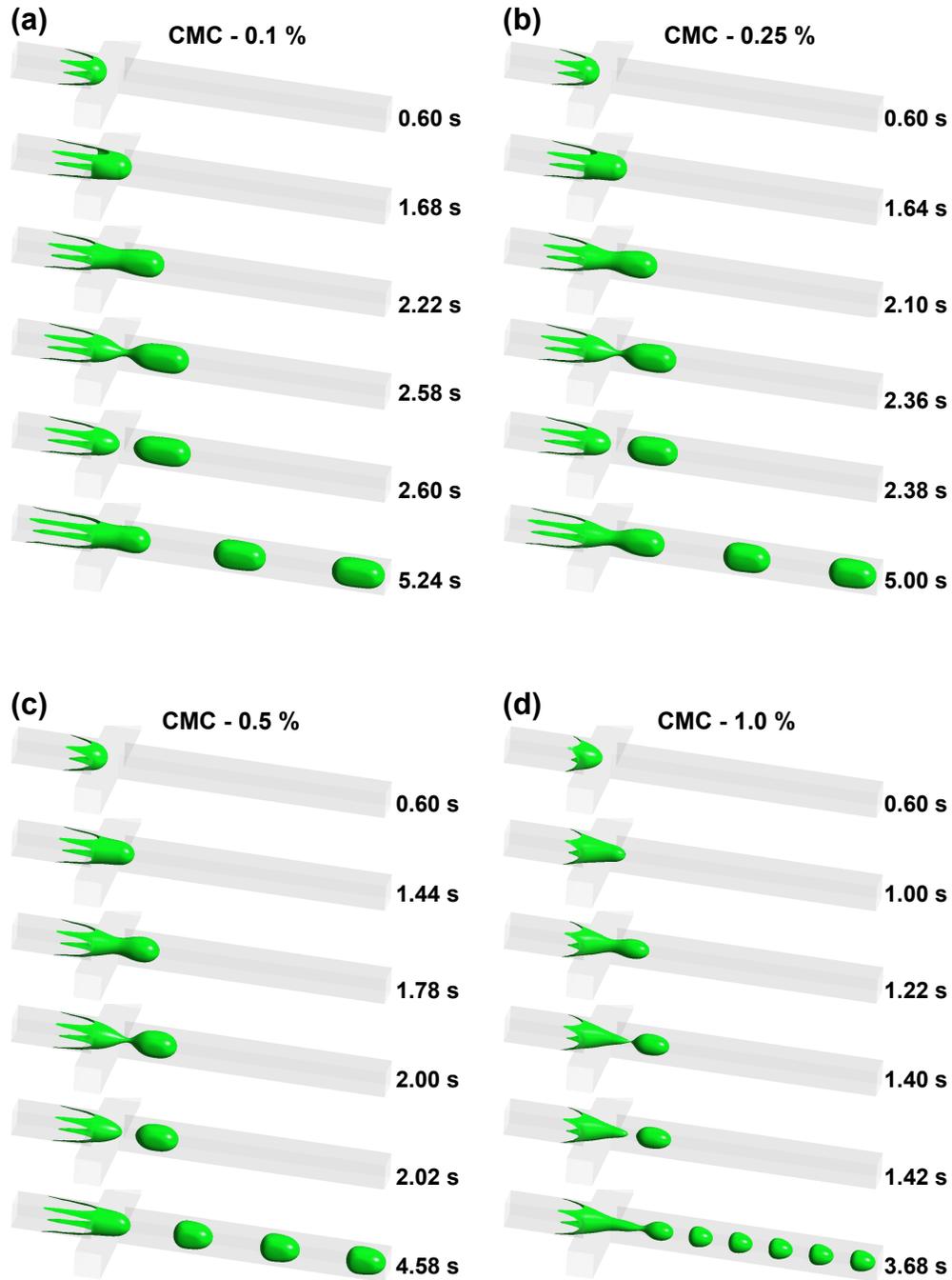} 
	\caption{Three-dimensional temporal evaluation on the effect of continuous phase concentration on droplet formation for  (a) CMC-0.1\%, (b) CMC-0.25\%, (c) CMC-0.5\%, and (d) CMC-1.0\% at $Q_{c} = 20 ~\mu L/min$ and $Q_{d} = 10 ~\mu L/min$.}
	\label{fig-4}
\end{figure*}

\begin{figure*}[!ht] 
	\centering
	\includegraphics[width=0.90\textwidth]{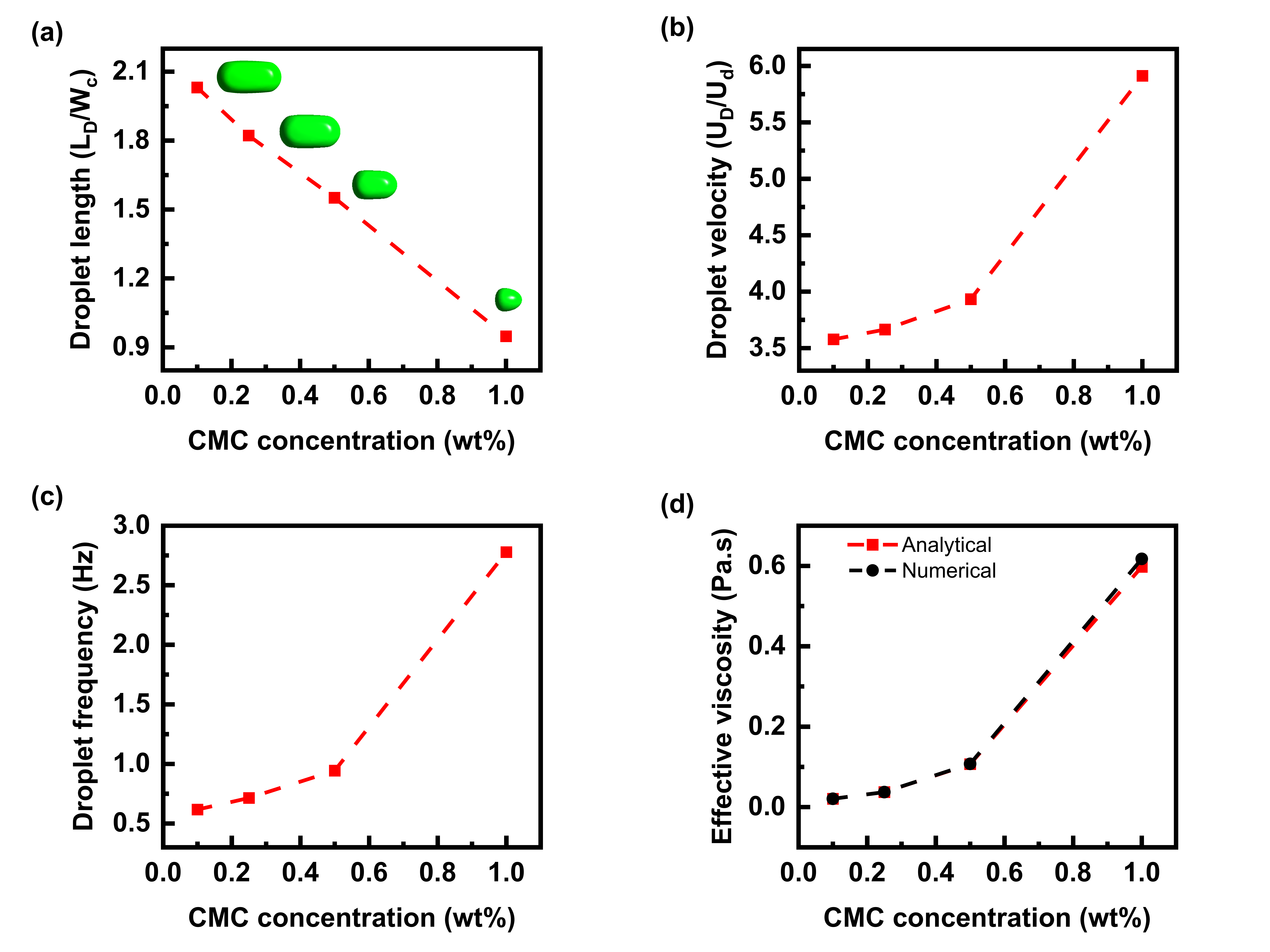} 
	\caption{ Effect of continuous phase concentration on (a) non-dimensional droplet length, (b) non-dimensional droplet velocity, (c) droplet formation frequency, and (d) comparison of simulated effective viscosity with analytical at $Q_{c} = 20 ~\mu L/min$ and $Q_{d} = 10 ~\mu L/min$.}
	\label{fig-6}
\end{figure*}

\begin{figure*}[] 
	\centering
	\includegraphics[width=0.80\textwidth]{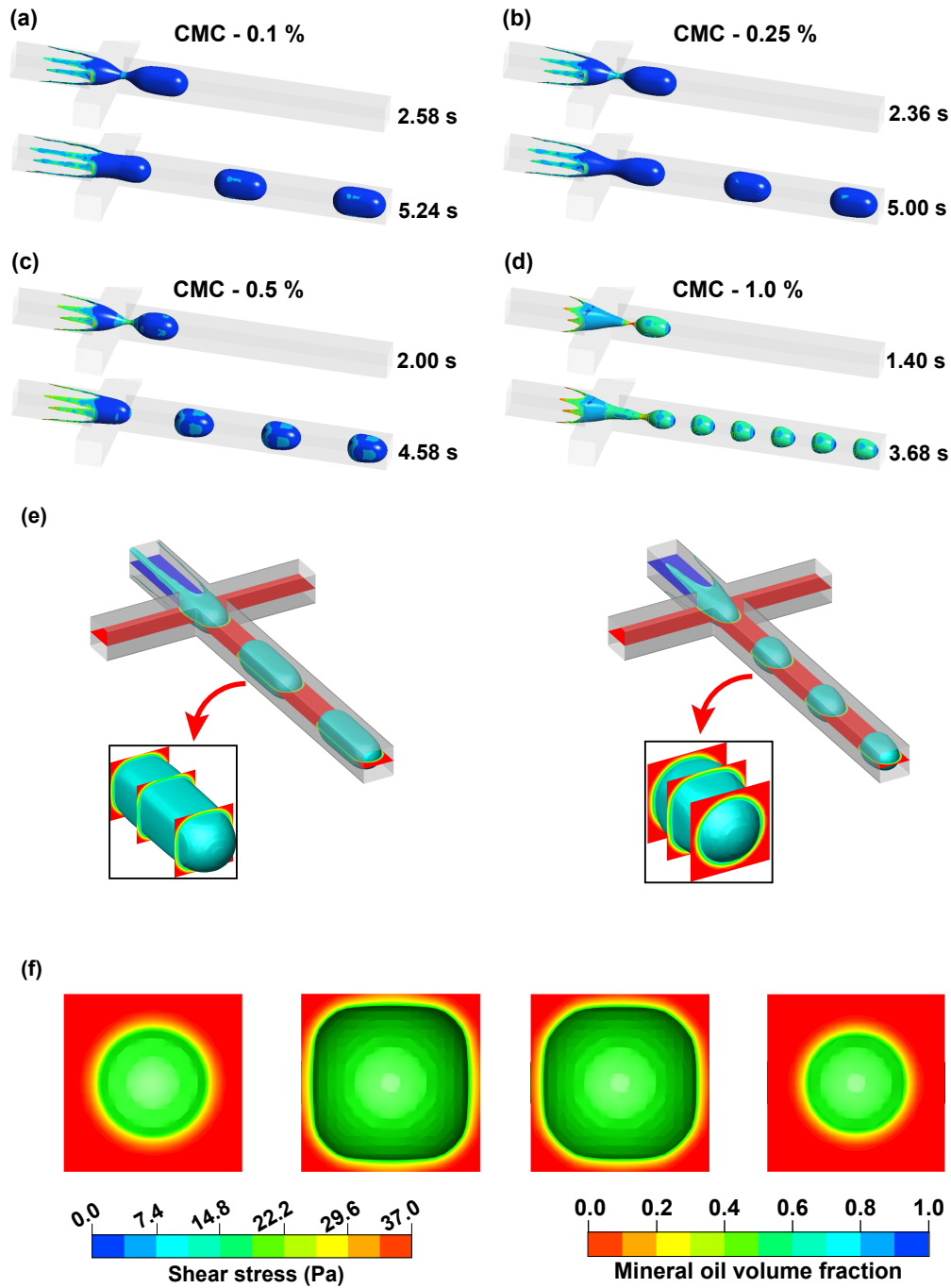} 
	\caption{ The effect of continuous phase concentration on droplet generation and shear stress magnitude for  (a) CMC-0.1\%, (b) CMC-0.25\%, (c) CMC-0.5\%, and (d) CMC-1.0\% at $Q_{c} = 20 ~\mu L/min$ and $Q_{d} = 10 ~\mu L/min$. (e) Three-dimensional iso-surface of the liquid film thickness in squeezing and dripping regime and (f) liquid film visualization around the droplet surrounded by the CMC-0.5\% solution at $Q_{c} = 10 ~\mu L/min$ and $Q_{d} = 10~\mu L/min$. }
	\label{fig-5}
\end{figure*}

\begin{figure*}[!ht] 
	\centering
	\includegraphics[width=0.80\textwidth]{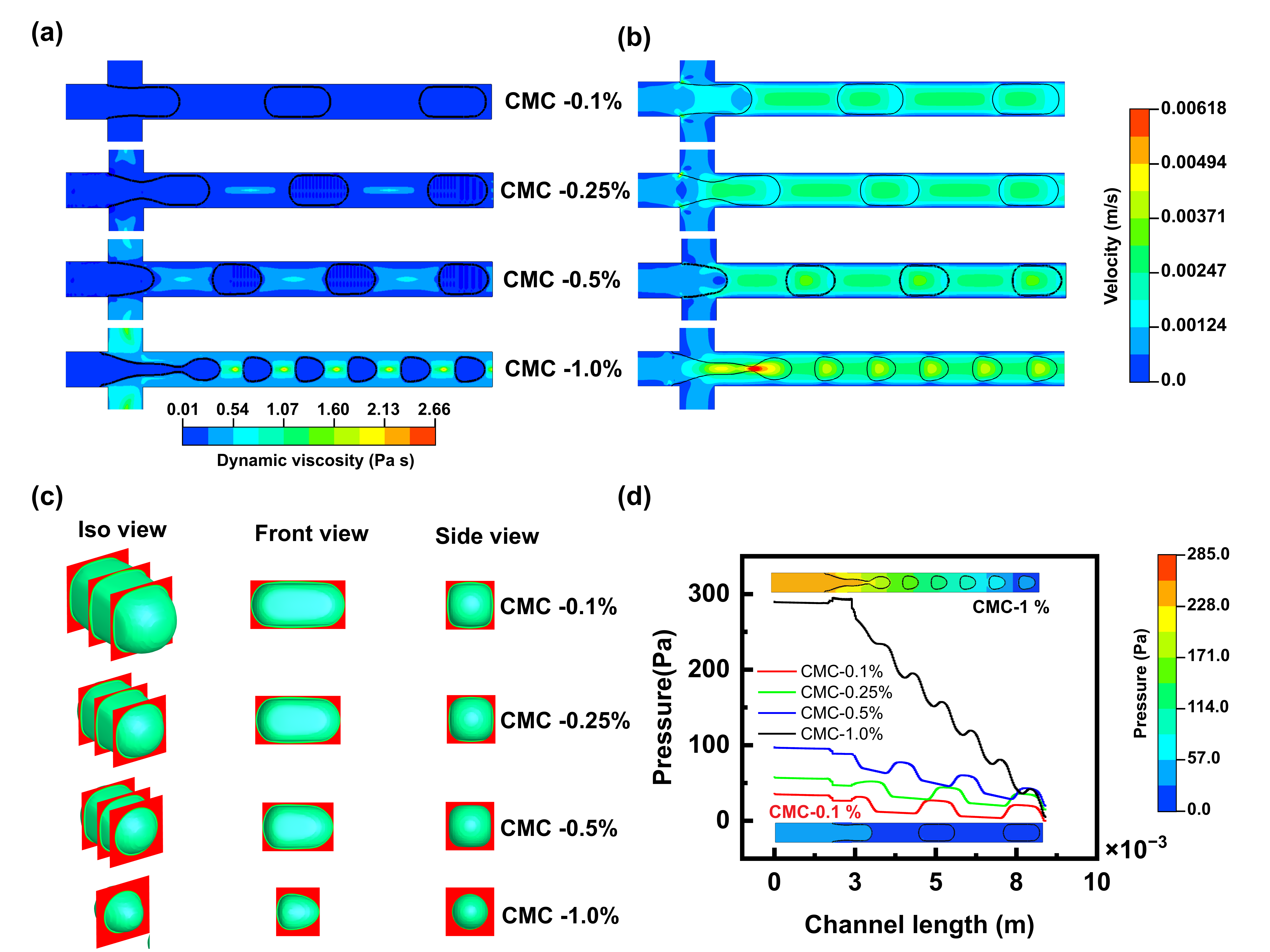} 
	\caption{ Effect of continuous phase concentration on (a)  dynamic viscosity, (b) flow field velocity, (c) liquid film thickness, and (d) pressure profile along the channel at $Q_{c} = 20 ~\mu L/min$ and $Q_{d} = 10 ~\mu L/min$.}
	\label{fig-7}
\end{figure*}

CMC solution, a shear-thinning non-Newtonian power-law fluid with concentrations CMC-0.1\%, CMC-0.25\%, CMC-0.5\%, and CMC-1.0\%, is used as a continuous phase. The viscosity of the CMC solution increases with increasing the concentration from 0.1\% to 1.0\%. The effective viscosity of the CMC solution is calculated analytically using the eq \ref{eqn11} \cite{rhodes2024introduction}.

\begin{equation}
\mu_{\text {eff }}=K\left(\frac{3 n+1}{4 n}\right)^n\left(\frac{8 v}{W_c}\right)^{n-1}
\label{eqn11}
\end{equation}

where $K$,~$n$,~$v$, and $W_{c}$ are the consistency index $(Pa.s^{n})$, power law index, velocity~(m/s), and width of the channel (m), respectively. At fixed fluid velocity and the microchannel width, the change in the CMC solution's concentration results in changes in viscosity (functions of $K$ and $n$). The hydrodynamic pressure, interfacial tension force, and shear forces are the three important forces that act on the dispersed phase and govern the formation of droplets \cite{agarwal2020dynamics}. The emerging dispersed phase (mineral oil)  into the microchannel experiences shear force and hydrodynamic pressure from the continuous phase. The interfacial tension resists the forces acting on the dispersed phase and minimizes the deformation. The upstream pressure buildup and continuous shear force overcome the interfacial tension force, eventually leading to the droplet formation. \\

The effect of continuous phase concentration on droplet formation is shown in Fig. \ref{fig-4}. Figure \ref{fig-4}a shows the time evolution of mineral oil droplet formation in a CMC-0.1\% concentration. At time 0.6 s, the mineral oil emerges near the junction, and the dispersed flow moves forward, reaching the main channel's inlet at time 1.68 s. The dispersed phase occupies the entire channel and squeezes into the main channel, observed at time 2.22 s. Due to the flow being blocked by the dispersed phase, there is pressure build-up from the upstream continuous phase on the dispersed phase, and this continuous pressure force and shear force acting on either side of the dispersed phase experiences the thinning phenomena, observed at 2.58 s. At time 2.60 s, the thinned dispersed phase eventually resulted in the pinch off from the bulk dispersed phase and formed as a droplet and stable droplet generation is shown at time 5.24 s.\\

From Fig. \ref{fig-4}a,b,c, and d, as the concentration of CMC solution increases, the shear force acting on the emerging dispersed phase increases, which results in the rate of thinning of dispersed phase thread increases, which is observed for CMC-0.1\%, CMC-0.25\%, CMC-0.5\%, and CMC-1.0\% at 2.22 s, 2.10 s, 1.78 s, and 1.22 s, respectively. The rate of droplet formation is also increased as the CMC concentration increases, as well as the thinning rate, as can be observed at 2.60 s, 2.38 s, 2.02 s, and 1.42 s for CMC-0.1\%, CMC-0.25\%, CMC-0.5\%, and CMC-1.0\%, respectively.\\

Figure \ref{fig-6}a shows the quantification of non-dimensional droplet length~($L_{D}/W_{C}$) with the effect of the CMC concentration. From Fig. \ref{fig-6}a, it is evident that at lower CMC concentration, the droplet length is two times the width ($W_{C}$) of the channel. As the concentration increases, the droplet size is reduced, and the percentage reduction of droplet size for CMC-0.1\% to CMC-0.25\% is 10.3\%, for CMC-0.25\% to CMC-0.5\% is 14.8\%, and for CMC-0.5\% to CMC-1.0\% is 38.8\% is the highest reduction in size is quantified due to the dominant viscous forces acting on dispersed phase. One of the important parameters to measure and quantify is the droplet velocity, which helps to understand the amount of time the droplet spends in the microchannel, especially fluids involving reactions. From Fig. \ref{fig-6}b, it is evident that the droplet velocity is increased with concentration, and the droplet velocity at CMC-1.0\% is significantly high due to the high shear force, which also results in droplet formation frequency shown in Fig. \ref{fig-6}c. The CMC solution's effective viscosity, calculated analytically using equation \ref{eqn11}, is compared with viscosity obtained from the simulation data shown in Fig. \ref{fig-6}d. Both analytical and simulation data are in good agreement, showing the CLSVOF model's accuracy. An experimental work of \citet{dong2023formation} reported a similar trend using the analytical correlation, and the numerical work of \citet{jammula2024numerical}  is in line with the current trend. Comparison of quantitative data with respect to CMC concentration is shown in Table S1.\\

\noindent As the concentration of CMC solution increases, the magnitude of the shear stress acting on the dispersed phase increases. The magnitude of shear stress is observed at 2.58 s, the pinch-off point as shown in Fig. \ref{fig-5}a. Furthermore, the magnitude of shear stress at the pinch-off point of the dispersed phase for CMC-0.25\% and CMC-0.5\% is observed from Fig. \ref{fig-5}b and c at 2.36 s and 2.00 s, respectively. For CMC-1.0\%, the shear stress induced on the dispersed phase at the pinch-off point is of the highest magnitude, which can be observed in Fig. \ref{fig-5}d at 1.40 s. Figure \ref{fig-5}e shows the three-dimensional droplet visualization with a mid-plane showing the liquid film occupied around the droplet in both squeezing and dripping regimes. Figure \ref{fig-5}f shows the planner view of the vertical slice of the droplet along with the liquid film (i.e., continuous phase) thickness around the droplet with the precise interface capturing using CLSVOF.\\

Qualitative visualization of dynamic viscosity and velocity in the channel is shown in Fig. \ref{fig-7} a and b. Figure \ref{fig-7}a shows the viscosity contour of CMC-0.1\% solution with a lower viscosity magnitude. For both the dispersed phase and continuous phase, the viscosity is almost equal. As the CMC concentration increases, the viscosity of the CMC solution also increases. However, the magnitude of the viscosity is highest at the center of the channel and minimal near the walls. This is because the shear rate near the walls is maximum, and the center of the channel is minimum. The region with the droplet shapes highlighted is the viscosity of the mineral oil, which is the same in all the contours from CMC-0.1\% to CMC-1.0\%. From Fig. \ref{fig-7}b, the velocity magnitude in the microchannel is qualitatively visualized with the effect of CMC concentration. As the concentration increases, the velocity magnitude also increases. In spite of this, due to the no-slip boundary condition, the magnitude of the velocity is almost zero near the walls and maximum in the center of the channel.\\

Liquid film thickness is the continuous phase fluid around the droplet and between the droplet and channel walls. Figure \ref{fig-7}c shows the liquid film occupied around the droplet, and as the CMC concentration increases, the droplet size is reduced, resulting in increased liquid film occupied around the droplet. Figure \ref{fig-7}d shows the pressure profile along the channel length for the four CMC concentrations. The pressure at the inlet of the channel increases as the concentration of CMC increases. However, the pressure gradually decreases to zero at the channel's exit due to pressure outlet boundary conditions. The pressure is increased as the concentration is increased due to the viscosity, which follows the Hagen-Poiseuille equation ($\Delta P \propto \mu $)\\
\begin{figure*}[] 
	\centering
	\includegraphics[width=\textwidth]{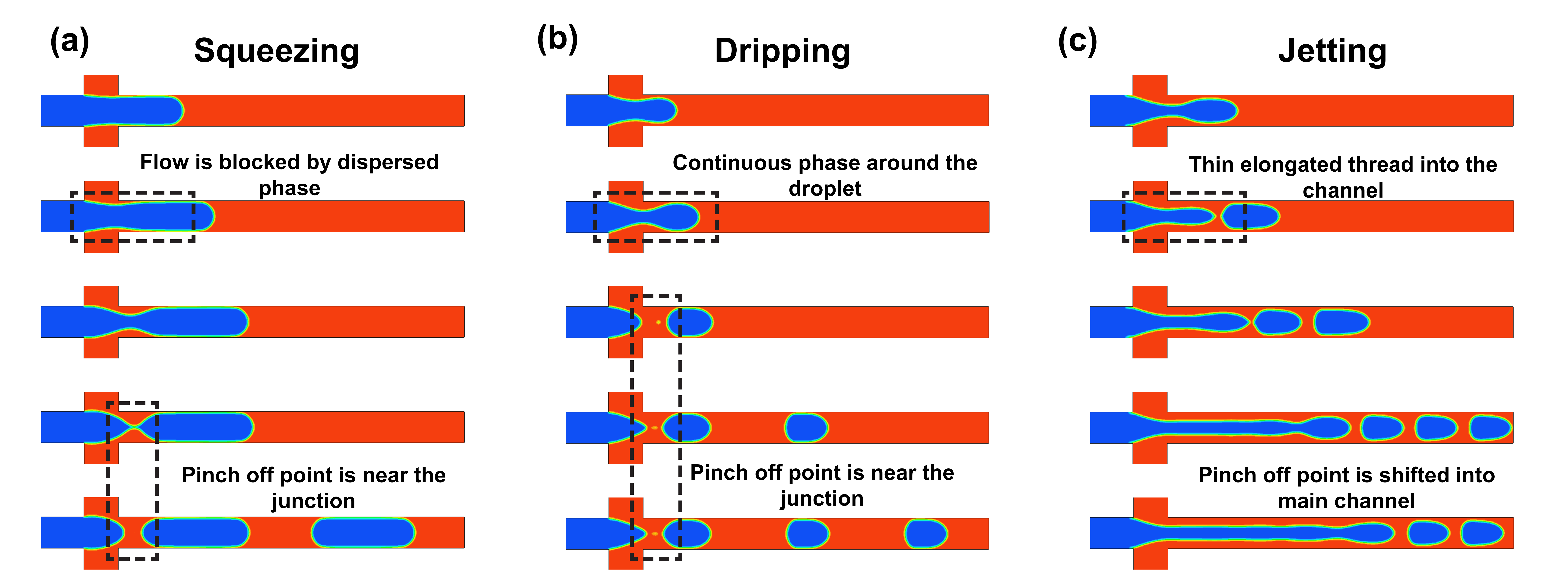} 
	\caption{ Flow regime  (a) Squeezing in CMC-0.1\% at $Q_{c} = 10 ~\mu L/min$ and $Q_{d} = 10 ~\mu L/min$. (b) Dripping in CMC-0.5\% at $Q_{c} = 30 ~\mu L/min$ and $Q_{d} = 10 ~\mu L/min$. (c) Jetting in CMC-1.0\% at $Q_{c} = 20 ~\mu L/min$ and $Q_{d} = 18 ~\mu L/min$.}
	\label{fig-s2}
\end{figure*}

\begin{figure*}[] 
	\centering
	\includegraphics[width=0.80\textwidth]{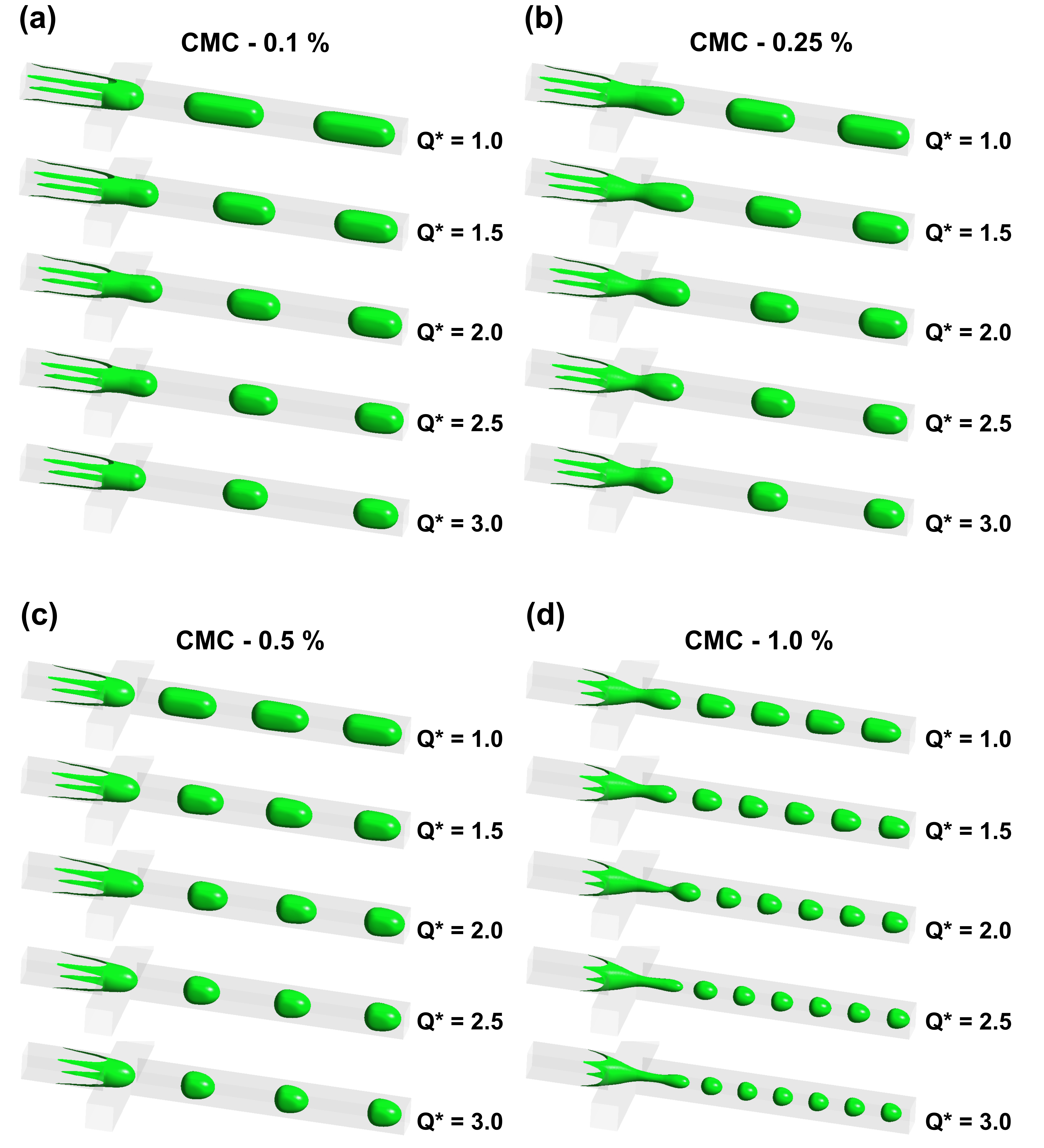} 
	\caption{ Effect of continuous phase flow rate on droplet size (a) CMC-0.1\%, (b) CMC-0.25\%, (c) CMC-0.5\%, and (d) CMC-1.0\% at $Q_{c} =10 ~\mu L/min $ to  $ 30 ~\mu L/min$ and $Q_{d} = 10 ~\mu L/min$.}
	\label{fig-8}
\end{figure*}

\begin{figure*}[] 
	\centering
	\includegraphics[width=\textwidth]{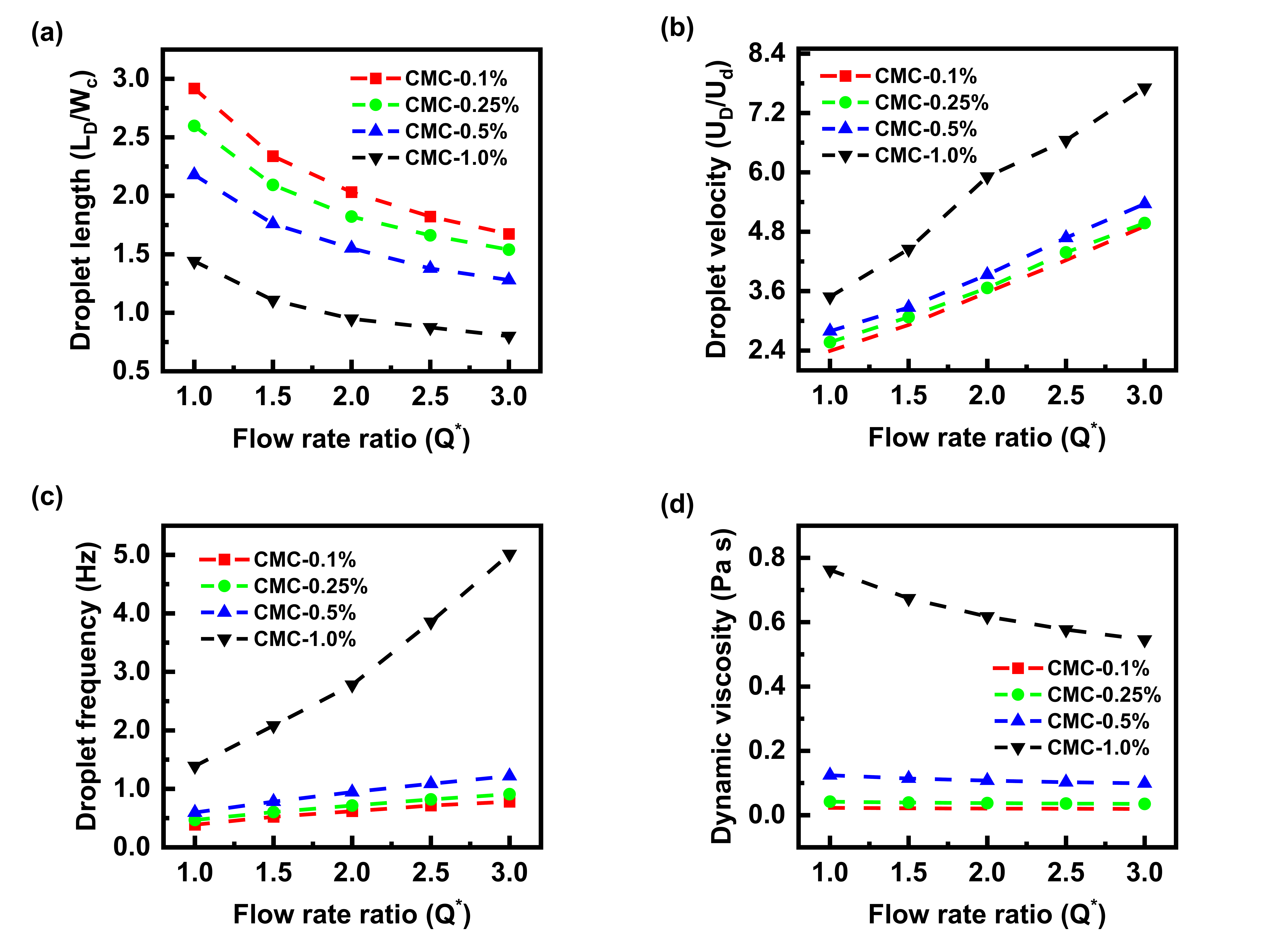} 
	\caption{ Effect of continuous phase flow rate on (a) non-dimensional droplet length, (b) non-dimensional velocity, (c) droplet formation frequency, and (d) dynamic viscosity at $Q_{c} =10 ~\mu L/min $ to  $ 30 ~\mu L/min$ and $Q_{d} = 10 ~\mu L/min$.}
	\label{fig-9}
\end{figure*}

 Flow regimes are the critical characterization of typical flow patterns and droplet formation. Three types of flow regimes are observed for the considered range of operating conditions such as squeezing, dripping, and jetting regime \cite{shang2017emerging}. The squeezing regime results in a pressure gradient due to the dispersed phase which occupies the entire channel and block the continuous upstream phase without any gap between the fluid interface and channel wall. As the pressure gradient increases beyond the dispersed phase's internal pressure, it forces the phase into the main channel, where necking occurs at the junction, leading to droplet formationThe droplets formed in the squeezing regime move downward in the microchannel with no gap between the interface and the channel wall, as shown in Fig. \ref{fig-s2}a. \\
 
In the dripping regime, the dispersed phase starts necking at the junction with a gap between the channel wall and interface. The droplet formation occurs within the junction limits shown in Fig. \ref{fig-s2}b. In the jetting regime, the dispersed phase experiences more inertial force and moves into the channel with a thin thread without pinch-off, and the droplet formation occurs inside the main channel far away from the microchannel junction shown in Fig. \ref{fig-s2}c. The droplet pinch-off position moves downward in the jetting regime, whereas the droplet pinch-off position is not changed in the squeezing and dripping regimes, as shown in Fig. \ref{fig-s2}.

\subsection{Effect of continuous phase flow rate}

\begin{figure*}[] 
	\centering
	\includegraphics[width=\textwidth]{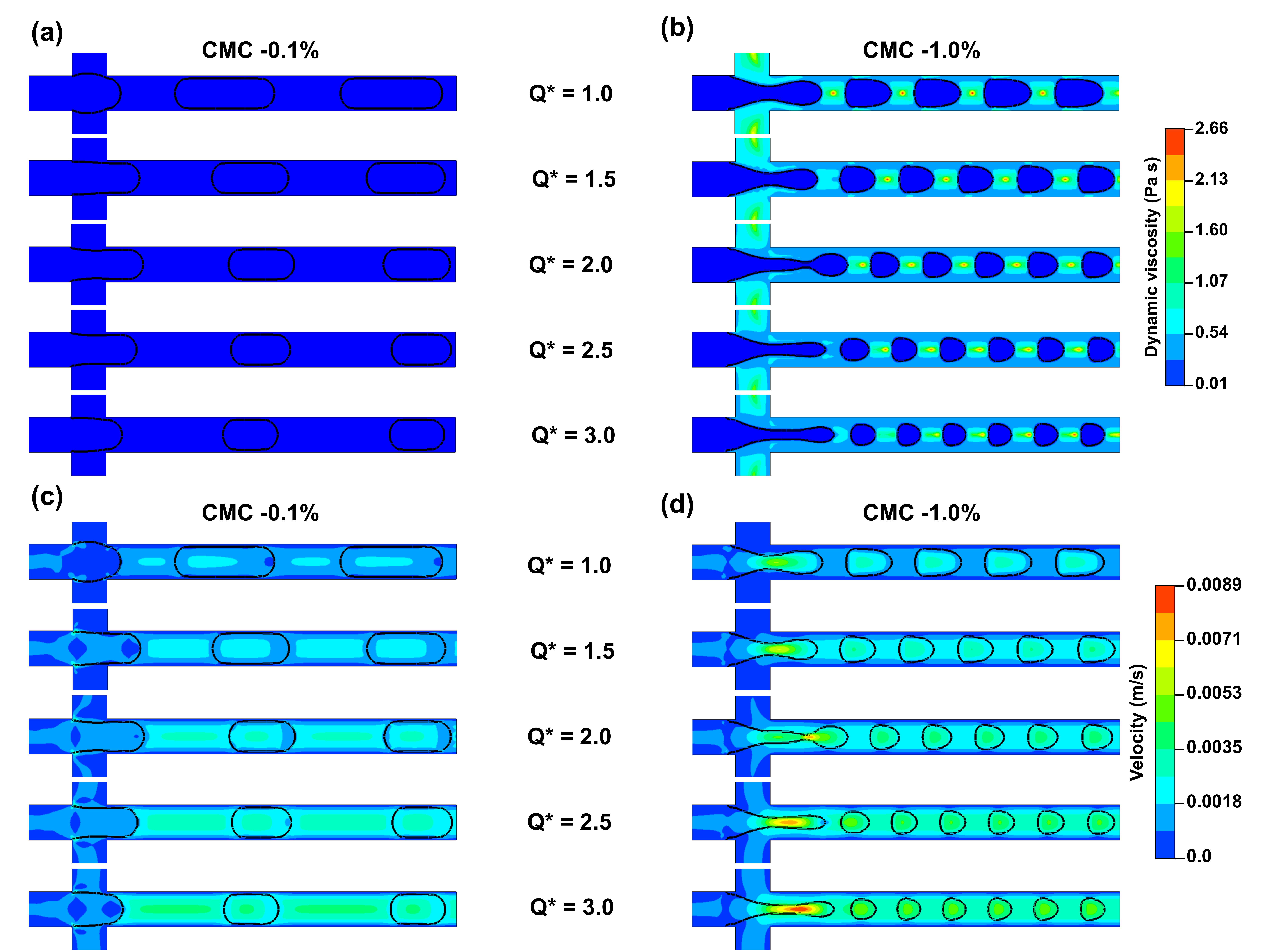} 
	\caption{ Effect of continuous phase flow rate on (a) dynamic viscosity for CMC-0.1\%, (b)  dynamic viscosity for CMC-1.0\%, (c) flow field velocity for CMC-0.1\%, and (d) flow field velocity for CMC-1.0\% at $Q_{c} =10 ~\mu L/min $ to  $ 30 ~\mu L/min$ and $Q_{d} = 10 ~\mu L/min$.}
	\label{fig-10}
\end{figure*}

\begin{figure*}[] 
	\centering
	\includegraphics[width=\textwidth]{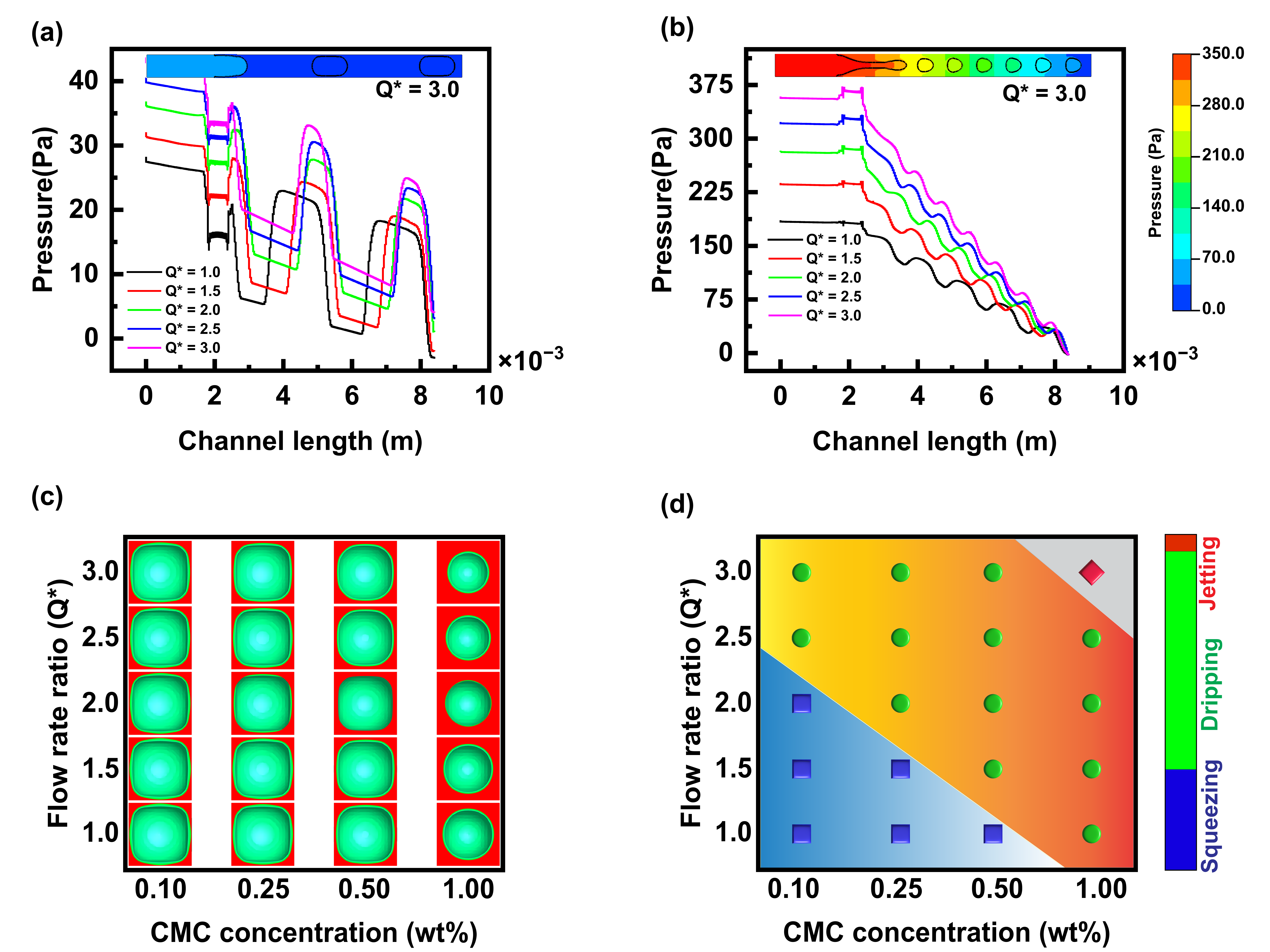} 
	\caption{ Effect of continuous phase flow rate on  (a) pressure profile along the channel length for CMC-0.1\%, (b) pressure profile along the channel length for CMC-1.0\%, (c) liquid film thickness, and (d) flow regime map at $Q_{c} =10 ~\mu L/min $ to  $ 30 ~\mu L/min$ and $Q_{d} = 10 ~\mu L/min$.}
	\label{fig-11}
\end{figure*}

\begin{figure*}[] 
	\centering
	\includegraphics[width=0.80\textwidth]{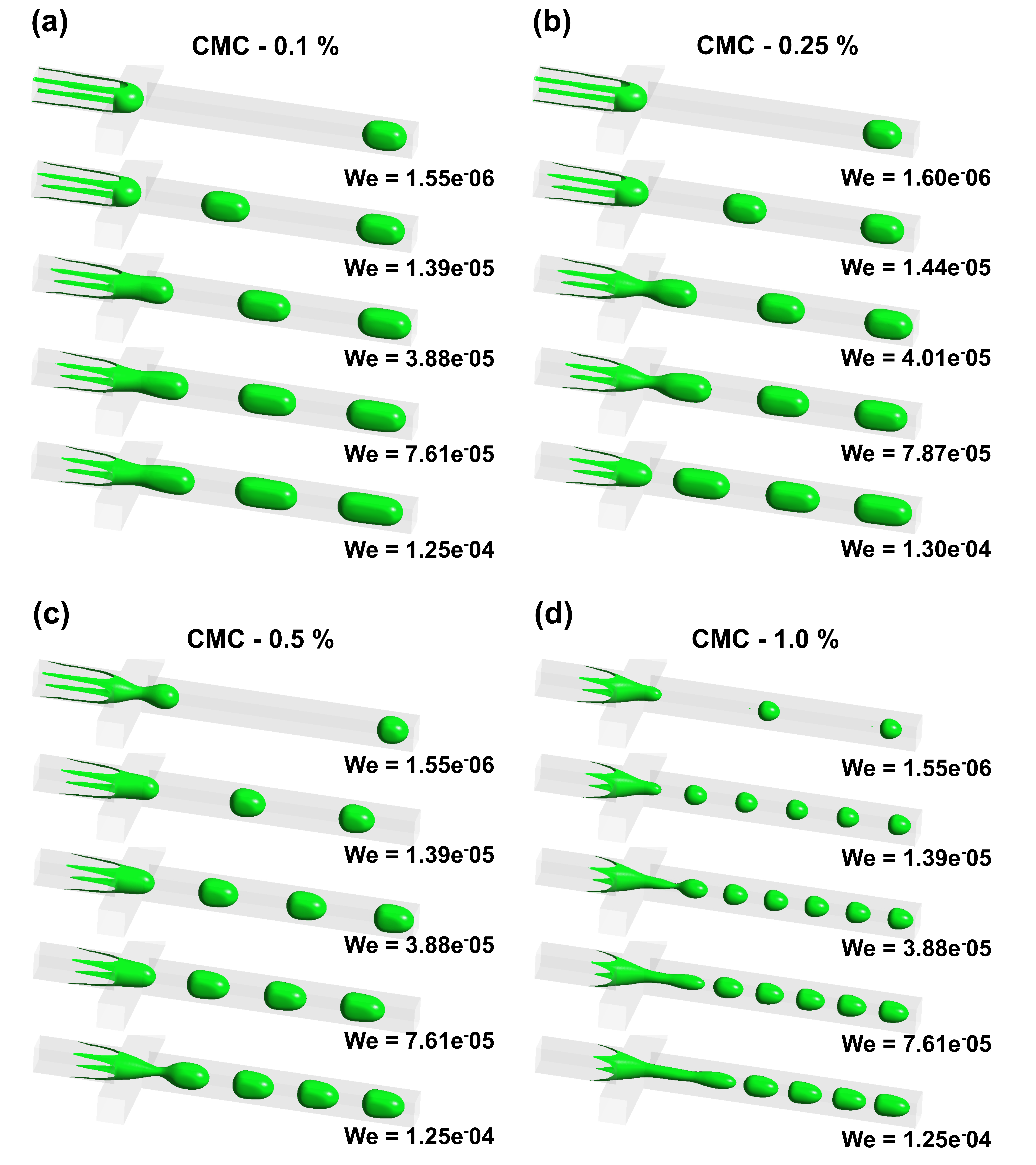} 
	\caption{ Effect of dispersed phase flow rate on droplet size (a) CMC-0.1\%, (b) CMC-0.25\%, (c) CMC-0.5\%, and (d) CMC-1.0\% at $Q_{c} = 20 ~\mu L/min$ and $Q_{d} = 02 ~\mu L/min $ to  $ 18 ~\mu L/min$.}
	\label{fig-12}
\end{figure*}

\noindent Apart from geometry and fluid properties, operating conditions have a significant impact on the ability to generate controlled droplets. This section systematically investigates the effect of continuous phase flow rate on droplet generation and their dynamics. Keeping the dispersed phase flow rate constant~($10 ~\mu L/min$), the continuous phase flow rates varied from $ 10 ~\mu L/min $ to  $ 30 ~\mu L/min$. Further in this section (Q* = $Q_{c}/Q_{d}$), a non-dimensional form of flow rate ratio is used to represent the flow rate, where $Q_{c}$ is continuous phase flow rate and $Q_{d}$ is dispersed phase flow rate.\\

Figure \ref{fig-8}a shows the qualitative visualization of droplet formation in CMC-0.1\% solution. At a low flow rate of the continuous phase (Q* =1.0), large plug-shaped droplets are observed. As the flow rate increases from Q* = 1.0 to Q* = 3.0, there is a noticeable decrease in the droplet length. A similar trend is observed in Fig. \ref{fig-8}b,c, and d. However, the droplet pinch-off position at higher CMC concentration is shifted from the junction to the downstream direction of the channel. The droplets formed in the lower CMC concentrations for all the flow rates are in plug shape, whereas for CMC-0.5\% at Q* = 2.5, Q* = 3.0, and for all the flow rates of CMC-1.0\%, the droplets are in cone shape with front end higher curvature.\\

To understand deeper insights into the influence of continuous phase flow rate, the droplet dynamics are quantified in Fig. \ref{fig-9}. Figure \ref{fig-9}a shows the non-dimensional droplet length ($L_{D}/W_{C}$) for all the CMC concentrations over a flow rate ratio Q* = 1.0 to 3.0 is quantified.
It is evident that the droplet size decreases with the increase in the flow rate for all the CMC concentrations due to the induced hydrodynamic pressure on the dispersed phase increasing with the flow rate of the continuous phase. However, the rate of droplet size decrease in the CMC-1.0\% solution is higher due to higher shear force.  \citet{chen2021pressure} attributed this to the increasing flow rate of the continuous phase generates smaller droplets. \\
 
 Non-dimensional droplet velocity ($U_{D}/U_{d}$) and formation frequency (Hz) with flow rate ratio (Q*) is increased for all the CMC concentrations due to the inertial forces inducing more hydrodynamic pressure on dispersed phase as shown in Fig. \ref{fig-9}b and c. However, the \% change in droplet velocity and formation frequency are 30\% and 75\%, respectively, from CMC-0.5\% to CMC-1.0\% at flow rate ratio Q* = 3.0 due to the high shear force. The dynamic viscosity of CMC solution, a shear-thinning fluid, will decrease as the shear rate increases. The shear rate is a function of the flow rate, and Fig.~\ref{fig-9}d shows the dynamic viscosity of the CMC solution. As the flow rate increases, the dynamic viscosity decreases for all the CMC concentrations but the decrease of the dynamic viscosity of lower CMC concentrations is insignificant, whereas for the higher concentrations, specifically for CMC-1.0\%, the decrease in dynamic viscosity is significant due to the as the concentration increases the shear thinning nature of the solution is increased. \\

 The qualitative visualization of the effect of continuous phase flow rate on dynamic viscosity and velocity within a microchannel is shown in Fig.~\ref{fig-10}. Figure \ref{fig-10}a shows the dynamic viscosity within the microchannel for CMC-0.1\% solution, and the dynamic viscosity of mineral oil and CMC-0.1\% are almost equal, and there is no significant change in the contours of the viscosity with increasing the flow rate from Q* = 1.0 to 3.0. Figure \ref{fig-10}b shows the dynamic viscosity of the mineral and CMC-1.0\% within the microchannel. The magnitude of viscosity in the droplet regions is constant in all the contours from Q* = 1.0 to 3.0 due to the mineral being a Newtonian fluid whose viscosity is constant. Whereas the viscosity other than droplet regions continuously decreases with the flow rate ratio from Q* = 1.0 to 3.0 due to CMC-1.0\% being a non-Newtonian shear-thinning fluid. However, the viscosity magnitude near the walls is minimum, and at the center, it is maximum due to it being vice versa to the shear rate.\\

 Figure \ref{fig-10}c and d show the velocity magnitudes within the microchannel with respect to the flow rate ratio (Q*) for CMC-0.1\% and CMC-1.0\% solutions, respectively. The velocity magnitude from the entrances of the dispersed phase is constant in all the contours, whereas the velocity magnitude from the inlet of the continuous phase is increased from Q*=1.0 to 3.0, however, the combined velocity magnitude within the microchannel from the junction where the two phases interact is increased with increasing the flow rate of continuous phase flow rate is shown in Fig. \ref{fig-10} c. A similar velocity magnitude to figure \ref{fig-10}c is observed at the entrances of the microchannel in Fig. \ref{fig-10}d, but the overall velocity magnitude is higher with flow rate ratio in CMC-1.0\% compared to the CMC-0.1\% solution due to the higher shear force contributing the velocity magnitude.\\

The pressure profile along the channel length for CMC-0.1\% solution over the flow rate ratio Q* =1.0 to 3.0 is shown in Fig. \ref{fig-11}a. As the flow rate increases, the inlet pressure continuously increases. However, the pressure at the outlet for all the flow rates is zero. The peaks of the pressure profile indicated the droplet's position within the microchannel, and the pressure peak is due to the axial and radial curvature of the droplet. Figure \ref{fig-11}b shows the pressure profile along the channel length for CMC-1.0\% solution over the flow rate ratio Q* =1.0 to 3.0. The pressure increases with the flow rate ratio, but the magnitude of the pressure is higher in CMC-1.0\% compared to CMC-0.1\%. This is due to the pressure is a function of viscosity and flow rate as per the Hagen Poisellue equation ($\Delta P \propto \mu  Q $).\\

Figure \ref{fig-11}c shows the liquid film thickness around the droplet for four CMC concentrations over the flow rate ratio Q* =1.0 to 3.0. For CMC concentrations 0.1\% and 0.25\%, the droplet length is decreased with flow rate. However, the droplet width is not significantly affected, and the liquid film thickness around the droplet looks similar for CMC-0.1\ and CMC-0.25\%. Whereas the liquid film thickness for the CMC concentrations 0.5\% and 1.0\% along with droplet length, the droplet width also significantly decreased with flow rate. Therefore, the liquid film occupied around the droplet increased as the flow rate increased. The flow regime map of droplet formation for all the CMC concentrations with respect to the flow rate ratio is developed and shown in Fig. \ref{fig-11}d. The flow regime of droplet formation is in the squeezing regime for CMC concentrations 0.1\%, 0.25\%, and 0.5\% at lower flow rates, and as the flow increases, the regime is shifted from squeezing to dripping from CMC-0.5\% to CMC-0.1\% in descending order. Whereas the flow regime of droplet formation for CMC-1.0\% solution is in the dripping regime for all the flow rates.\\

\subsection{Effect of dispersed phase flow rate}

\begin{figure*}[] 
	\centering
	\includegraphics[width=\textwidth]{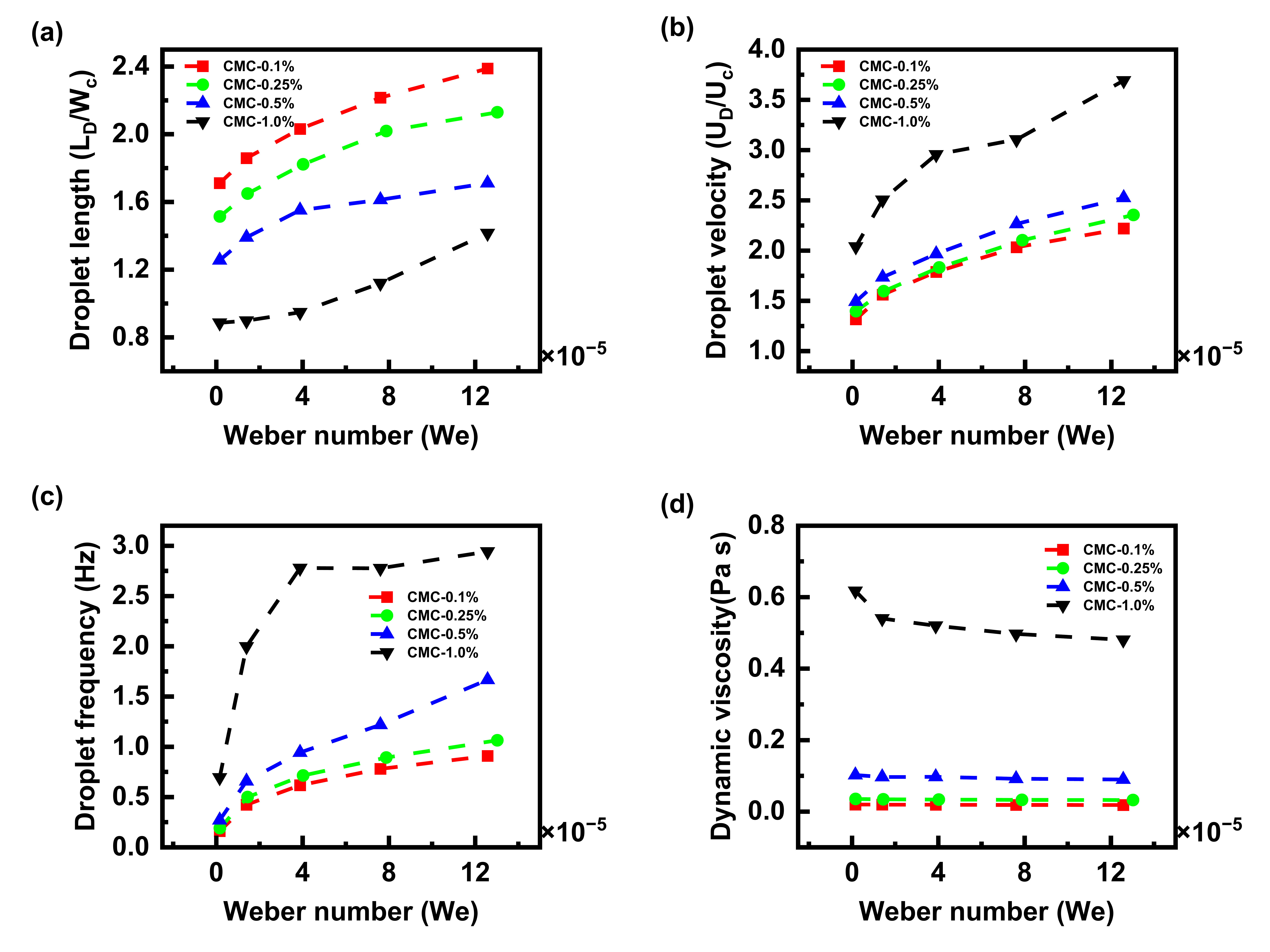} 
	\caption{  Effect of dispersed phase flow rate on (a) non-dimensional droplet length, (b) non-dimensional velocity, (c) droplet formation frequency, and (d) dynamic viscosity at $Q_{c} = 20 ~\mu L/min$ and $Q_{d} = 02 ~\mu L/min $ to  $ 18 ~\mu L/min$.}
	\label{fig-13}
\end{figure*}

\begin{figure*}[!ht] 
	\centering
	\includegraphics[width=\textwidth]{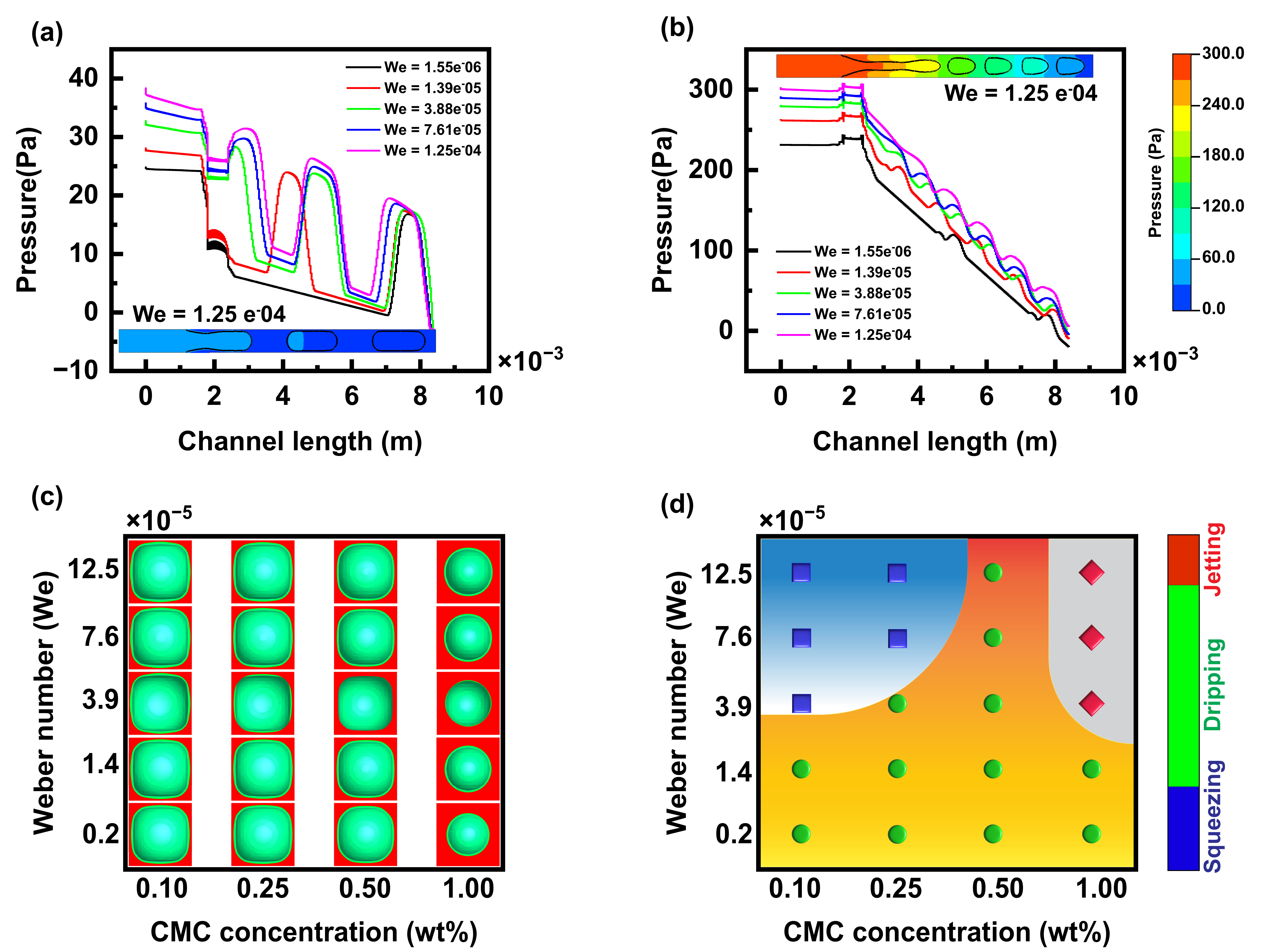} 
	\caption{ Effect of dispersed phase flow rate on   (a) pressure profile along the channel length for CMC-0.1\%, (b) pressure profile along the channel length for CMC-1.0\%, (c) liquid film thickness, and (d) flow regime map at $Q_{c} = 20 ~\mu L/min$ and $Q_{d} = 02 ~\mu L/min $ to  $ 18 ~\mu L/min$.}
	\label{fig-14}
\end{figure*}
\noindent This section investigates the influence of the dispersed phase flow rate on the droplet size, velocity, and formation frequency. The continuous phase flow rate is constant in this section, and the dispersed phase flow rate is operated over a range from $2 ~\mu L/min $ to  $ 18 ~\mu L/min$. The dispersed phase flow rate in this section is represented as Weber number, which is the ratio of inertial force to the interfacial tension force ($\mathrm{We}=\frac{\rho u_{d}^2 w_{c}}{\sigma}$). The generation of mineral oil droplets with varying flow rates of mineral oil surrounded by CMC solutions is shown in figure \ref{fig-12}. When the dispersed phase enters the microchannel, the shear force from the continuous phase acts on the dispersed phase and pinches off into the droplet. As the dispersed flow rate increases, the volume of the dispersed phase entering the channel per unit time increases and the inertial force from the dispersed phase flow rate resists the shear force from the continuous phase. Until the shear force is sufficient enough to pinch off the dispersed phase into the droplets, the dispersed phase volume increases, and the droplet size increases. \\

Figure \ref{fig-12}a shows the droplet generation in CMC-0.1\% solution from $We = 1.55e^{-06} ~ to ~ 1.25e^{-04}$, at lower flow rate the droplet is smaller in size and a single droplet is observed in the microchannel. With the increase in flow rate, both droplet formation rate and size increased. Figure \ref{fig-12}b shows the droplet formation in CMC-0.25\% solution over a range of $We = 1.60e^{-06} ~ to ~ 1.30e^{-04}$. Similar to the CMC-0.1\% solution, a droplet formation rate with an increased droplet size trend is observed in CMC-0.25\% with an increasing flow rate. However, the droplet size is smaller, and formation frequency is higher in CMC-0.25\% solution due to the shear force acting on the dispersed phase is increased. Figure \ref{fig-12}c shows the droplet formation in CMC-0.5\%, the droplet size and formation rate increases with increasing the dispersed phase flow rate. The droplet in CMC-0.5\% is observed to be cone-shaped, whereas, in CMC-0.1\% and CMC-0.25\%, the droplets are plug-shaped. The droplet formation and size increased in CMC-1.0\%. However, the position of the droplet pinch-off moved in the downward direction into the microchannel as the flow rate of the dispersed phase increased, shown in Fig. \ref{fig-12}d. This is attributed to the dominant shear force, which is unlikely in lower CMC concentrations.\\

The quantification of droplet length, droplet velocity, frequency, and dynamic viscosity of CMC solution with the influence of the dispersed flow rate are shown in Fig. \ref{fig-13}.  When the dispersed phase enters the microchannel, it tries to spread in all directions, and for lower CMC concentrations (0.1\%, 0.25\%), the shear force acting on the dispersed phase is minimal. Therefore, the dispersed phase grows radially before expanding in the axial direction, and when the dispersed phase flow rate increases, the fluid thread grows in the axial direction due to the space constraint in the radial direction. However, for higher CMC concentrations (0.5\%, 1.0\%), the shear force is sufficient enough to resist the dispersed phase to grow radially, and in all CMC concentrations with increasing the flow rate of the dispersed phase, the Weber number increases, and the non-dimensional droplet length ($L_{D}/W_{C}$) with respect to the Weber number is increasing as shown in Fig. \ref{fig-13}a. With increasing the flow rate of the dispersed phase, the inertial force dominates within the microchannel, resulting in an increase in droplet velocity and formation frequency. The non-dimensional droplet velocity ($U_{D}/U_{c}$)  is shown in Fig. \ref{fig-13}b, and droplet formation frequency (Hz) is shown in Fig. \ref{fig-13}c. The dynamic viscosity of the CMC solution is shown in Fig. \ref{fig-13}d. The effect of the Weber number on the dynamic viscosity of lower CMC concentrations is insignificant. However, for CMC-1.0\%, the dynamic viscosity decreases with the Weber number due to the CMC-1.0\% solution having a higher shear thinning nature than the lower CMC concentrations.\\

The qualitative visualization of viscosity and velocity magnitude in a microchannel with the Weber number for the lower CMC (0.1\%) and higher CMC (1.0\%) concentrations are shown in Fig. S2. Figure S2 a, shows the viscosity magnitude inside a microchannel for CMC-0.1\% over the Weber number $We = 1.55 \times 10^{-06}$ ~ to ~ $1.25 \times 10^{-04}$. The viscosity magnitude of mineral oil is constant irrespective of the flow rate due to the Newtonian nature. In contrast, the CMC-0.1\% is non-Newtonian, but the viscosity of the CMC-0.1\% solution is nearly close to the viscosity of the mineral oil. The viscosity magnitude inside a microchannel for CMC-1.0\% by varying the Weber number is shown in Fig. S2 b. The viscosity magnitude of droplet regions is mineral oil, and outside the droplet regions are CMC-1.0\% solution. The viscosity magnitude difference from the droplet region to outside the droplet region is clearly noticeable as the CMC-1.0\% high viscous solution, whereas the viscosity of the mineral oil is low. As the flow rate of the dispersed phase increases, the viscosity of the magnitude within the channel is reduced from $We = 1.55$ $\times 10^{-06}$ ~ to ~ $1.25 \times 10^{-04}$.  The velocity magnitude within the microchannel for CMC -0.1\% and CMC-1.0\% solutions with dispersed phase flow rate is shown in Fig. S2 c and d, respectively. The velocity magnitude within the microchannel increases with increasing the dispersed phase flow rate for both cases. However, the velocity magnitude for CMC-1.0\% is higher due to the higher shear forces also influencing the velocity magnitude.\\

The pressure profile along the channel length with the influence of the dispersed phase flow rate for all the CMC concentrations is analyzed, and for CMC concentrations, 0.1\% and 1.0\% are shown in Fig. \ref{fig-14}a and b, respectively. Figure \ref{fig-14}a shows the pressure profile along the channel length for CMC-0.1\% over $We = 1.55 \times 10^{-06} ~ to ~ 1.25 \times 10^{-04}$. As the Weber number increases, the inlet pressure increases due to the inertial force. However, due to the pressure outlet boundary condition, the pressure at channel length $8\times10^{-03}$ m is zero. At lower CMC concentrations, as the droplet forms, it occupies the entire channel width and blocks the flow, causing the local pressure to rise, shown as the peak. As the droplet moves downward, where the flow is established with a continuous phase, the pressure drop is observed as a valley. Figure \ref{fig-14}b shows the pressure profile along the channel length for a CMC concentration 1.0\% over $We = 1.55 \times 10^{-06} ~ to ~ 1.25 \times 10^{-04}$. Similar to the CMC-0.1\%, as the Weber number increases due to the inertial force, the inlet pressure increases. However, the magnitude of the pressure is higher for CMC-0.1\% due to the pressure rise contributing from the viscous and inertial forces. Unlike the peak and valley, the pressure profile is periodic in the CMC-1.0\% due to the droplets formed not completely occupying the channel, and the flow is steady with a continuous phase around the droplet. The liquid film around the droplet for CMC concentrations with the influence of the dispersed phase flow rate is shown in Fig. \ref{fig-14}c. At lower CMC concentrations, the droplet size almost occupies the width of the channel as the flow rate increases. However, at higher CMC concentrations, significant liquid film thickness is observed due to the influence of shear force dominating the dispersed phase. As the Weber number increases, the liquid film thickness is reduced due to the inertial force at fixed shear force. The influence of the dispersed phase flow rate on the flow regime is shown in Fig. \ref{fig-14}d. For CMC-0.1\% at a low Weber number due to the balance between the shear force and interfacial tension, the droplets generated are uniform and in the dripping regime. As the Weber number increases, the excess inertial force acts against the shear force from the CMC-0.1\% transitioned into the squeezing regime. A similar trend is observed for a CMC-0.25\% concentration, and with increasing the Weber number, the transition of dripping to squeezing flow regime in droplet formation is observed. Dripping regime is observed for CMC-0.5\% concentration over the $We = 1.55 \times 10^{-06} ~ to ~ 1.25 \times 10^{-04}$ due to the balance between the shear force and interfacial tension forces of the dispersed and continuous phases. For CMC-1.0\% concentration, At a lower Weber number, the flow regime observed is dripping, and as the Weber number increases, the flow regime transits to the jetting regime due to the dominating shear and inertial forces.

\subsection{Effect of interfacial tension}

\begin{figure*}[] 
	\centering
	\includegraphics[width=\textwidth]{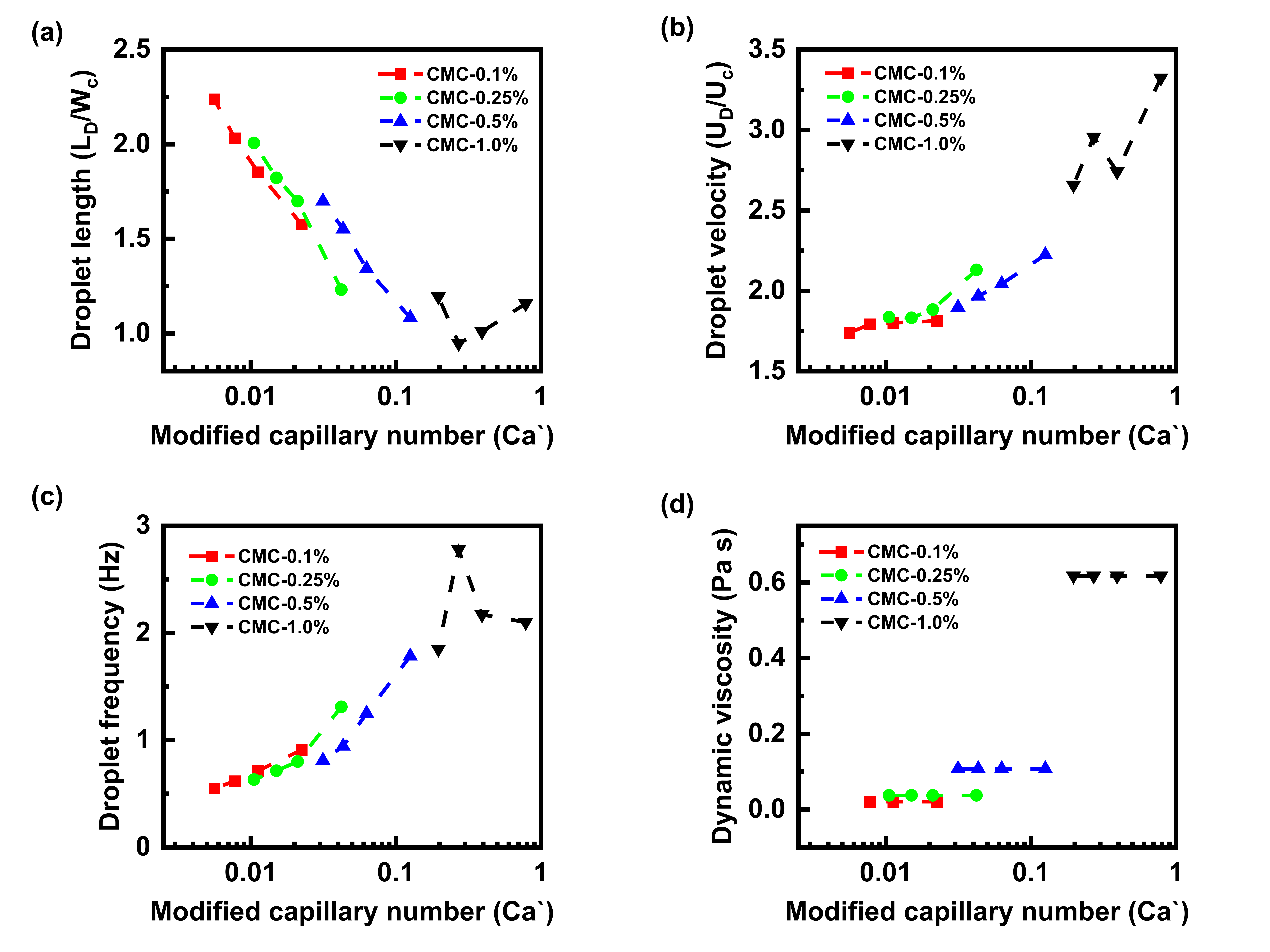} 
	\caption{ Effect of interfacial tension force on (a) non-dimensional droplet length, (b) non-dimensional velocity, (c) droplet formation frequency, and (d) dynamic viscosity at $Q_{c} = 20 ~\mu L/min$ and $Q_{d} = 10 ~\mu L/min $ .}
	\label{fig-15}
\end{figure*}

\begin{figure*}[] 
	\centering
	\includegraphics[width=\textwidth]{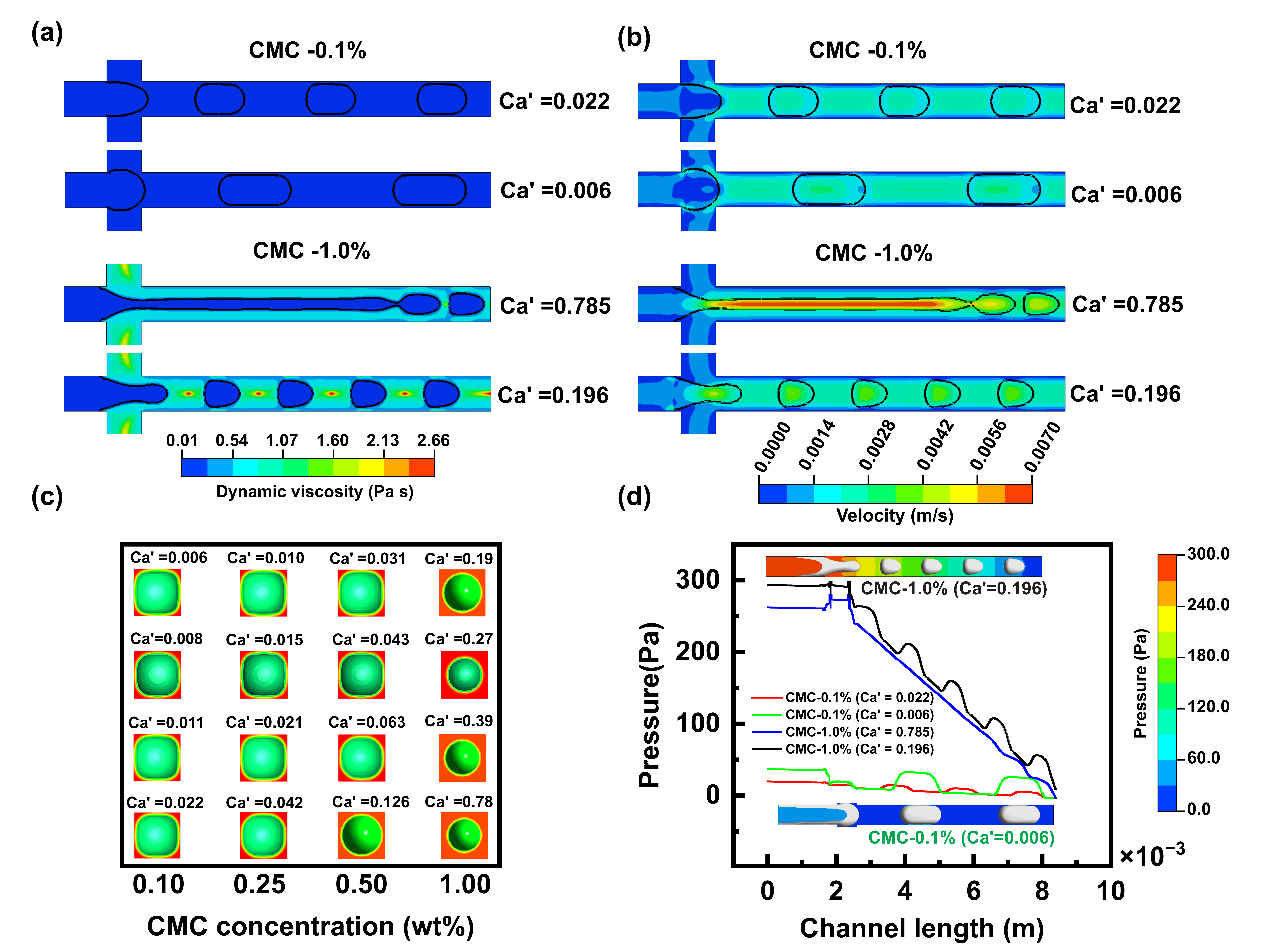} 
	\caption{  Effect of interfacial tension force on (a)  dynamic viscosity, (b)flow field velocity, (c) liquid film thickness, and (d) pressure profile along the channel for CMC-1\% at $Q_{c} = 20 ~\mu L/min$ and $Q_{d} = 10 ~\mu L/min$.}
	\label{fig-16}
\end{figure*}
\noindent One of the important forces that significantly influence droplet generation and droplet dynamics is the interfacial tension force between the two immiscible continuous and dispersed phases. The manipulation of interfacial tension using a surfactant in practical applications such as biomedical diagnosis, drug discovery, and chemical analysis plays a significant effect and provides insightful information for designing multiphase systems and microfluidic devices. This section systematically investigates the influence of interfacial tension on droplet formation, size, velocity, formation frequency, and pressure profile. 

In this work, interfacial tension is represented in terms of the Modified capillary number (Ratio of viscous forces to the interfacial tension force, ($Ca' = \frac{K v^{n}{W_{C}}^{n-1}}{\sigma}$). At a fixed flow rate of the continuous phase and dispersed phase is $Q_{c} = 20 ~\mu L/min$ and $Q_{d} = 10 ~\mu L/min $, respectively, the interfacial tension force is varied from 0.001 to 0.004 N/m and the corresponding modified capillary number ranging from 0.006 to 0.785 for all the CMC solutions. When the dispersed phase enters the microchannel due to the immiscibility with the continuous phase, it establishes an interface between the two immiscible fluids. The force acting on the interface is the interfacial tension force, which minimizes the surface area. When the dispersed phases advance into the channel due to the inertial force of the dispersed phase, the shear force and hydrodynamic pressure from the continuous phase act on the dispersed phase and stretch the dispersed phase into thinning. The interfacial tension acts opposite to shear force and hydrodynamic force and pulls back the dispersed phase inward, and at the maximum thinning position, the dispersed phase thread pinches off into the droplet.\\

Figure S3 shows the droplet formation mechanism in four CMC concentrations with the effect of modified capillary number, and with increasing the interfacial tension, the modified capillary number decreases $(Ca' \propto 1/\sigma)$. Figure S3a shows the droplet formation in CMC-0.1\% solution, as the interfacial tension force increases, at lower shear force, the dispersed phase thread squeezes into the main channel, and due to the almost equaling the width of the microchannel, the dispersed phase spreads downward of the microchannel leads into an increase in droplet length with a squeezing regime. Whereas, for CMC-0.25\%, the droplet formation at lower interfacial tension (higher $Ca'$) is in the dripping regime due to the increased shear force, as shown in Fig. S3 b.  Figure S3c shows the droplet formation mechanism in CMC-0.5\% solution at a low interfacial tension with an increased shear force acting on the dispersed phase. The droplet formation occurred downstream, far away from the junction, with an elongated thread. As the interfacial tension increases, the droplet pinch-off position shifts backward near the junction with an increase in droplet size. In CMC-1.0\%, at lower interfacial tension, the droplet formation is observed at the exit of the channel, and as the interfacial tension increases ($Ca'$ reduces), the droplet breakup from the dispersed thread shifts backward with an increase in droplet size, as shown in the Fig. S3 d. 

The quantification of droplet length, droplet velocity, formation frequency, and dynamic viscosity with the influence of the interfacial tension is shown in Fig. \ref{fig-15}. As the interfacial tension decreases, the modified capillary number increases and the non-dimensional droplet length ($L_{D}/W_{C}$) decreases with a modified capillary number for all the CMC concentrations of 0.1\%, 0.25\%, and 0.5\% shown in Fig. \ref{fig-15}a. However, for CMC-1.0\% at lower interfacial tension (higher $Ca'$), the droplet length is found to be higher than the droplet length at higher interfacial tension. This is due to the condition at lower interfacial tension, the shear force acting on the interface is more significant, which caused the dispersed phase to form a thin elongated thread before pinching off without expanding radially, the droplet pinched off from the dispersed phase is thin with higher in length. The interfacial tension force at the interface between two fluids holds the dispersed phase from the pinch-off, and as the interfacial tension force decreases, the shear force dominates over the interfacial tension force, hindering the droplet from growing before the pinch-off. At lower interfacial tension, the modified capillary number is higher, where the influence of shear force plays a significant role in the droplet pinch-off and droplet velocity. Figure \ref{fig-15}b shows the droplet velocity, as the interfacial tension decreases, the corresponding modified capillary number increases, and this results in droplet velocity increase, and at higher CMC concentration (CMC-1.0\%), the magnitude of droplet velocity is higher due to the dominant shear force. Figure \ref{fig-15}c shows the droplet formation frequency, increasing the modified capillary number increased the formation frequency for all the CMC solutions. However, the magnitude of the frequency is higher for CMC-1.0\%. The dynamic viscosity of the CMC solution with the influence of interfacial tension force is shown in Fig. \ref{fig-15}d. At a fixed flow rate of the continuous and dispersed phases, the influence of the interfacial tension on the dynamic viscosity of CMC solutions is not much affected due to the constant shear rate.\\

The qualitative visualization of the influence of the interfacial tension on the dynamic viscosity and velocity is shown in Fig. \ref{fig-16}a and b. Figure \ref{fig-16}a shows the dynamic viscosity magnitude within the microchannel for CMC-0.1\% and CMC-1.0\% continuous phase for the corresponding lower and higher interfacial tension force (represented as $Ca'$). The dynamic viscosity magnitude of CMC-0.1\% - mineral and CMC-1.0\% - mineral oil at their corresponding modified capillary numbers from $Ca'$ = 0.006 to 0.022 and $Ca'$ = 0.196 to 0.785, respectively, is not affected due to the constant shear and inertial force. Similar to Fig. \ref{fig-16}a, the velocity magnitude within the microchannel for CMC-0.1\% and CMC-1.0\% is not much affected however, for both CMC concentrations at higher Ca' (lower interfacial tension), the velocity magnitude is noticeably higher.  From Fig. S3, it is evident for all the cases, as the interfacial tension force increases, the droplet size increases irrespective of the droplet pinch-off position. When the droplet size is increased, the volume occupied by the droplet in a channel increases, and the liquid film between the droplet and channel wall reduces, and the liquid film thickness around the droplet within the microchannel is shown in Fig. \ref{fig-16}c. A similar trend of influence of interfacial tension force is represented in terms of the capillary number reported in the recent works of \citet{deepak2024film}. 

Figure \ref{fig-16}d shows the pressure profile along the channel length for CMC-0.1\% and 1.0\% for a lower and higher $Ca'$ with the influence of interfacial tension force. A continuous pressure drop was observed in all the cases. However, the pressure difference for CMC-1.0\% is higher due to the higher viscosity. Besides the viscosity, the interfacial tension contributes to the pressure profile, for a CMC-0.1\% at higher interfacial tension (lower $Ca'$), the pressure drop profile from the channel inlet to the exit is periodic and stable. A peak and valley profile is observed for CMC-0.1\% at higher interfacial tension force. There is a noticeable pressure increase as the interfacial tension increases ($Ca'$ decreases) due to the Laplace pressure $\Delta P=\sigma\left(\frac{1}{R_1}+\frac{1}{R_2}\right)$, and the peaks and valleys in a pressure profile is a result of the axial and radial curvatures of the droplet.

\section{CONCLUSION}
\noindent In this work, a 3D CFD simulation of droplet generation in a flow-focusing microchannel was conducted using a CLSVOF method for interface capturing. The rheological effects of shear thinning non-Newtonian fluid (CMC solution) on mineral oil droplet production were systematically investigated. The rheological parameters $K$ and $n$ were expressed in terms of viscosity of CMC concentrations 0.1\%, 0.25\%, 0.5\%, and 1.0\%. An in-depth numerical analysis of the effect of rheological parameters, flow rates of continuous and dispersed phases, and interfacial tension forces on droplet production and dynamics was explained. The qualitative and quantitative droplet dynamics, including droplet length, droplet velocity, formation frequency, liquid film thickness, pressure profile, and flow regimes for all the conditions numerically examined and reported. \\

From the findings, it is evident that the droplet length is decreased by increasing the concentration of CMC solution (i.e., varying the rheological parameters $K$ and $ 
n$) and flow rate of the CMC solution. Whereas the droplet length increased by increasing the dispersed phase flow rate and interfacial tension force. At lower CMC concentrations, the generated plug-shaped droplets showed characteristics of the squeezing regime, fully occupying the channel width. The CMC concentrations of 0.25\% and 0.5\% exhibited droplet formation in a dripping regime with a liquid film around the droplets at higher flow rates of the continuous and dispersed phases. At higher CMC concentrations, the droplets generated were in a dripping regime at lower flow rates and in the jetting regime at higher flow rates of the continuous and dispersed phases. The droplet velocity, formation frequency, liquid film thickness, and pressure profile increased significantly with higher CMC concentration and increased flow rates of the continuous and dispersed phases. To analyze droplet dynamics, the non-dimensional Weber number and modified Capillary number were expressed as functions of the dispersed phase flow rate and interfacial tension force. The fundamental understanding of the impact of non-Newtonian fluid on droplet generation in a flow-focusing microchannel contributes to advancements in the field of microfluidics.\\

Expanding on these findings, the results hold significant relevance for practical applications such as drug delivery, material synthesis, and various microfluidic technologies that require precise control over droplet size and shape. The capability to regulate droplet formation dynamics in shear-thinning fluids can be leveraged to improve processes in microfluidic systems. Future studies could examine the influence of different non-Newtonian fluid models with varying rheological properties and validate the numerical predictions with experimental observations. Furthermore, exploring droplet control in more intricate geometries or under dynamic flow conditions would strengthen the real-world applicability of this research.

\section*{ACKNOWLEDGEMENT}
\noindent 
This work is supported by the IIT Dharwad through the Seed Grant and Networking Fund (SGNF) scheme.

\section*{SUPPLEMENTARY MATERIAL}
See the supplementary material for FIG. \ref{fig-s1} ( comparison of simulation results of viscosity as a function of strain rate with experimental data), FIG. \ref{fig-s2} (flow chart of CLSVOF algorithm), TABLE \ref{tabS1} (Summary of effect of CMC concentration on droplet formation and dynamics at $Q_{c} = 20 ~\mu L/min$ and $Q_{d} = 10 ~\mu L/min$.), FIG. \ref{fig-s3}(qualitative analysis of dynamic viscosity and velocity with the effect of dispersed phase flow rate), and FIG. \ref{fig-s4} (qualitative representation of droplet formation with the effect of interfacial tension force).
\begin{figure*}[]
\centering
	\includegraphics[width=\textwidth]{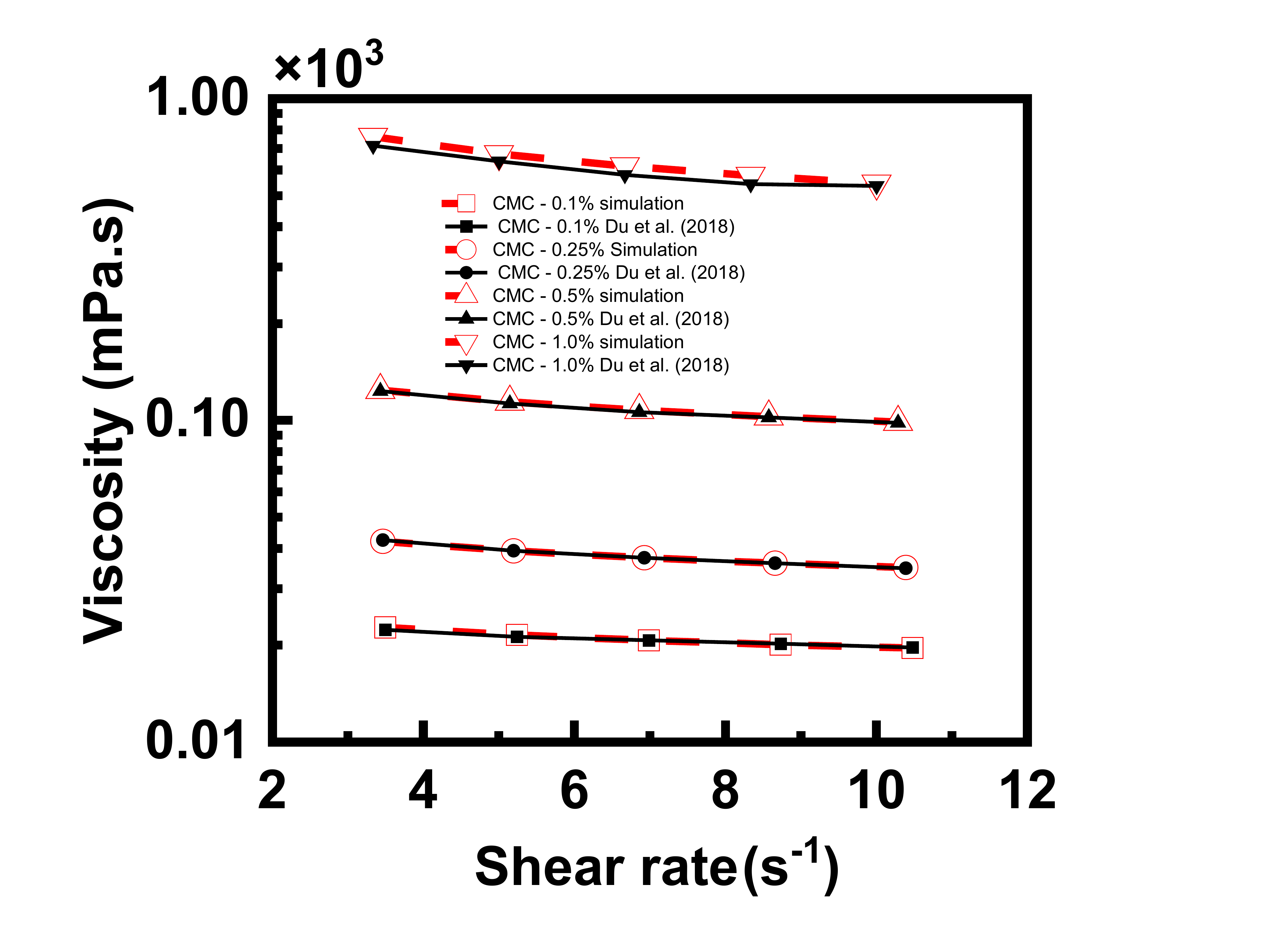} 
  \renewcommand{\thefigure}{S\arabic{figure}}
  \setcounter{figure}{0}
	\caption{ Comparision of simulation results of viscosity as a function of shear rate with \citet{du2018breakup} data.}
	\label{fig-s1}
\end{figure*}

\begin{figure*}[] 
	\centering
	\includegraphics[width=0.5\textwidth]{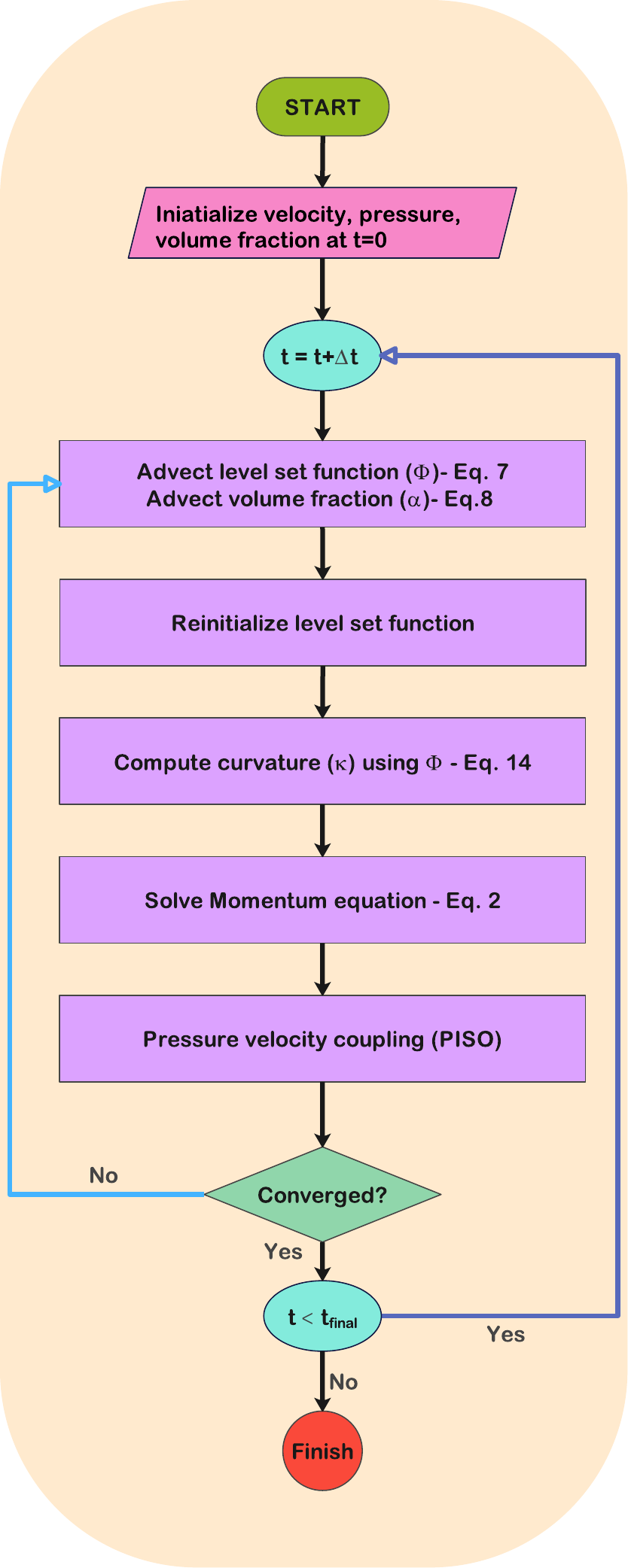} 
  \renewcommand{\thefigure}{S\arabic{figure}}
  \setcounter{figure}{1}
	\caption{ Flow chart of CLSVOF algorithm.}
	\label{fig-s2}
\end{figure*}

\renewcommand{\thetable}{S\arabic{table}}
\setcounter{table}{0}

\begin{table*}[!t]
\centering
\begin{tabular}{|c|c|c|c|c|}
\hline

\makecell{CMC\\concentration} & \makecell{Effective\\Viscosity (Pa.s)} & \makecell{Droplet\\length ($L_{D}/W_{C}$)} & \makecell{Droplet\\velocity ($U_{D}/U_{d}$)} & \makecell{Droplet\\frequency (Hz)}\\
\hline
CMC-0.1\% & 0.0207 & 2.0312 & 3.5770 & 0.62 \\ 
\hline
CMC-0.25\% & 0.0374 & 1.8219 & 3.665 & 0.7144 \\ 
\hline
CMC-0.5\% & 0.1075 & 1.551& 3.933& 0.9434 \\ 
\hline
CMC-1.0\% & 0.617 & 0.9478 & 5.9123 & 2.778 \\ 
\hline

\end{tabular}
\caption{Summary of effect of CMC concentration on droplet formation and dynamics  at $Q_{c} = 20 ~\mu L/min$ and $Q_{d} = 10 ~\mu L/min$.}
\label{tabS1}
\end{table*}

\begin{figure*}[] 
	\centering
	\includegraphics[width=\textwidth]{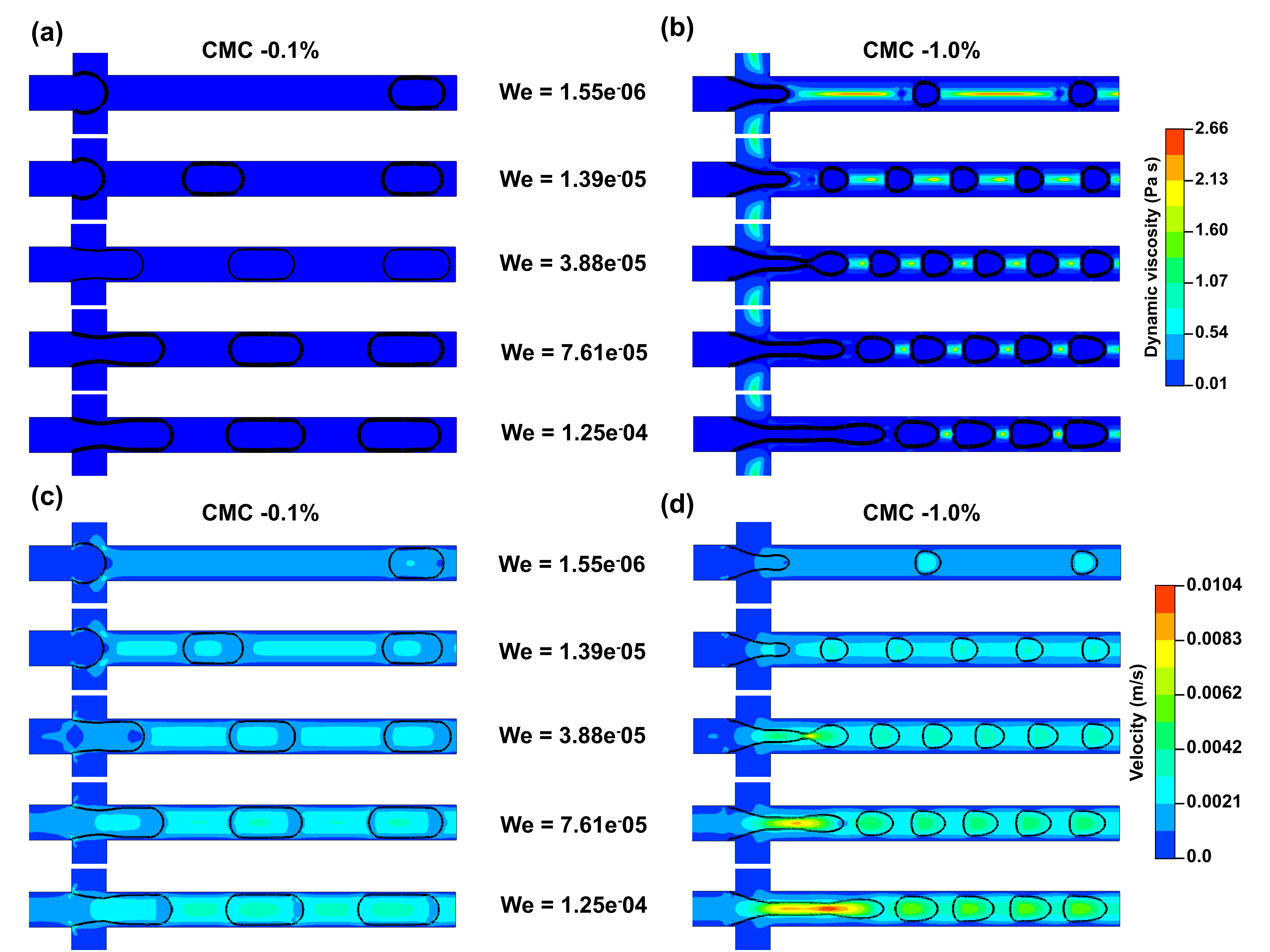} 
  \renewcommand{\thefigure}{S\arabic{figure}}
  \setcounter{figure}{2}
	\caption{Effect of dispersed phase flow rate on (a) magnitude of dynamic viscosity for CMC-0.1\%, (b) magnitude of dynamic viscosity for CMC-1.0\%, (c) magnitude of velocity for CMC-0.1\%, and (d) magnitude of velocity for CMC-1.0\% at $Q_{c} = 20 ~\mu L/min$ and $Q_{d} = 02 ~\mu L/min $ to  $ 18 ~\mu L/min$.}
	\label{fig-s3}
\end{figure*}

\begin{figure*}[] 
	\centering
	\includegraphics[width=\textwidth]{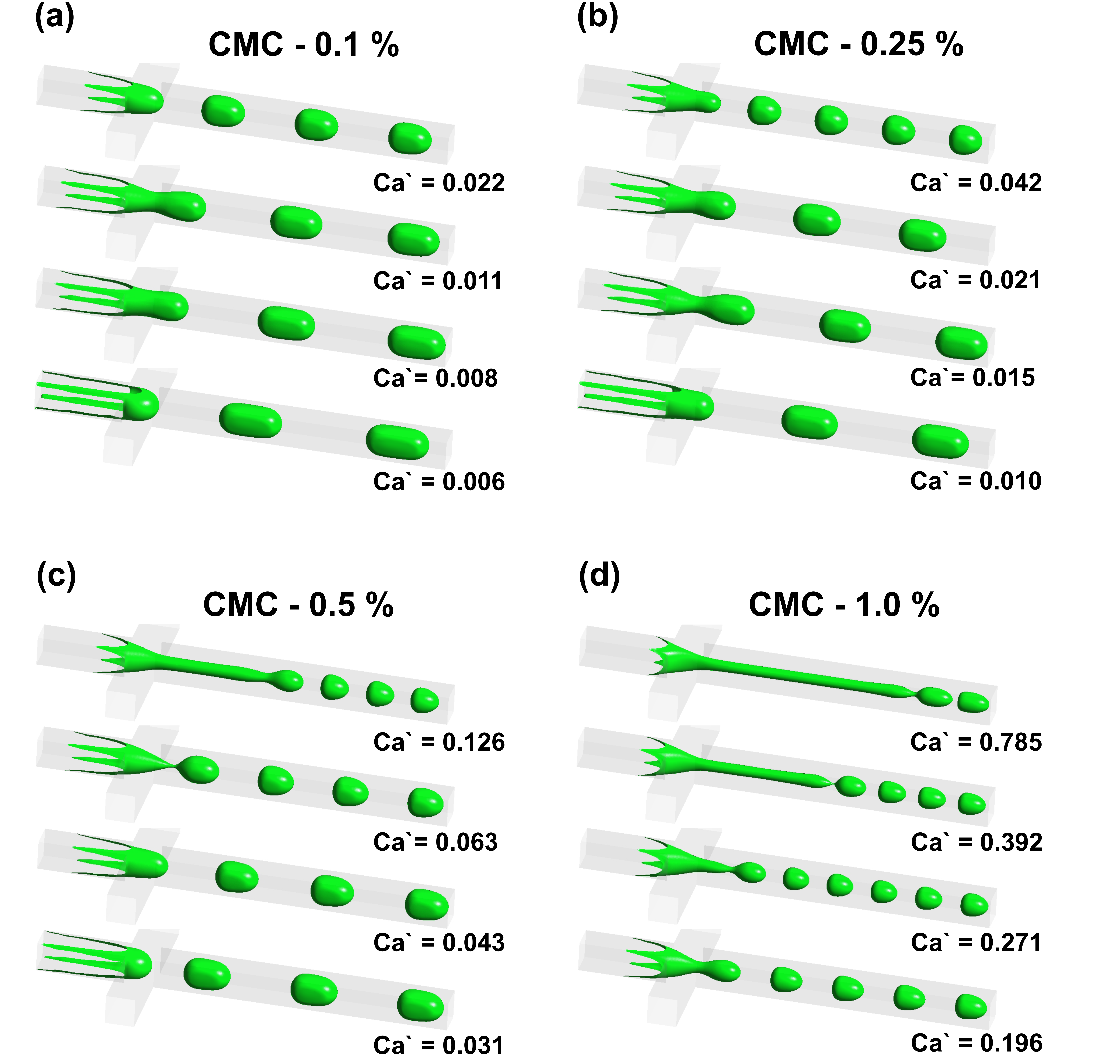} 
  \renewcommand{\thefigure}{S\arabic{figure}}
  \setcounter{figure}{3}
	\caption{ Effect of interfacial tension force on droplet size (a) CMC-0.1\%, (b) CMC-0.25\%, (c) CMC-0.5\%, and (d) CMC-1.0\% at $Q_{c} = 20 ~\mu L/min$ and $Q_{d} = 10 ~\mu L/min$.}
	\label{fig-s4}
\end{figure*}
\clearpage
\section*{AUTHOR DECLARATIONS}
\subsection*{Conflict of Interest}
The authors have no conflicts to disclose
\subsection*{Author Contributions}
\textbf{M. Jammula:} conceptualization (lead); methodology (lead); planned and performed the simulations (lead); model validation (lead); formal analysis (lead); software (lead); visualizations (lead);writing-original draft (lead); writing-review and editing(lead). \textbf{S. G. Sontti:} conceptualization (lead); methodology(supporting); data interpretation (supporting); project administration (lead); writing-review and editing (lead); resources(lead); and supervision (lead).

\section*{DATA AVAILABILITY}
\noindent The data supporting this study's findings are available from the corresponding author upon reasonable request.

\section*{NOMENCLATURE}
\begin{longtable} {l p{12cm}}
$B_{o}$ & Bond number ($\Delta \rho g W_{C}^2 / \sigma$)\\
$W_{e}$ & Weber number ($\frac{\rho u^2 W_{C}}{\sigma}$)\\
$C_{a}^{\prime}$ & modified Capillary number ($\frac{K u^n W_{C}^{(1-n)}}{\sigma} $)\\
$\vec{g}$ & gravitational force (m/$s^{2}$)\\
$W_{C}$ &  channel width (m) \\
$L_D$ & droplet length(m)\\
$H_D$ & droplet height (m)\\
$U_{D}$ & velocity of the droplet (m/s)\\
$U_{d}$ & velocity of the dispersed phase (m/s)\\
$U_{c}$ & velocity of the continuous phase (m/s)\\
$K$ &  consistency index (Pa.$s^n$)\\
$n$ & power law index\\
$t$ & time (sec)\\
$Q_c$ & flow rate of the continuous phase ($m^{3}/s$)\\
$Q_d$ & flow rate of dispersed phase ($m^{3}/s$)\\
$Q^{*}$ & ratio of flow rates ($Q_c/Q_d$)\\
$\vec{u}$ & velocity vector(m/s)\\
$p$ & pressure (Pa)\\
$\vec{F_{\sigma}}$ & interfacial tension force (N/m)\\
$\vec{\zeta}$ & position vector\\
$H(\Phi)$ & Heaviside function\\
$\omega$ & interface thickness (m)\\
$\delta(\Phi)$ & Dirac delta function\\
$\mu_{eff}$ & effective viscosity\\

\end{longtable}
\section*{Greek symbols}
\begin{longtable}{l p{12cm}}

$\mu$ & dynamic viscosity (Pa.s)\\
$\dot{\gamma}$ & shear rate (1/s)\\
$\rho$ & density (kg/$m^{3}$)\\
$\sigma$ & interfacial tension (N/m)\\
$\Phi$ & level set function\\
$\kappa$ & radius of curvature\\
$\theta$ & contact angle ($^{\circ}$)\\
$\alpha$ & volume fraction\\
$\gamma$ & distance from the interface to the bulk fluid

\end{longtable}
\section*{Subscripts}
\begin{longtable}{l p{12cm}}

$C$ & Channel\\
$D$ & droplet\\
$c$ & continuous phase\\
$d$ & dispersed phase\\

\end{longtable}
\nocite{*}
\bibliography{aipsamp}

\end{document}